\documentclass[12pt,aps,nofootinbib,superscriptaddress,preprintnumbers]{revtex4}

\usepackage{amsmath,amssymb,bm,hyperref}
\usepackage{subfigure}
\usepackage{float}
\usepackage{color}
\usepackage{graphicx}
\newcommand{\beq}{\begin{eqnarray}}
\newcommand{\eeq}{\end{eqnarray}}
\renewcommand\d{\partial}

\begin{document}
 \title{Instabilities in Anisotropic Chiral Plasmas}
\author{Avdhesh Kumar$^1$, Jitesh R. Bhatt}
\affiliation{Theoretical Physics Division, Physical Research Laboratory,
Navrangpura, Ahmedabad 380009 India}
\email{avdhesh@prl.res.in, jeet@prl.res.in}
\author{P. K. Kaw}
\affiliation{Institute for Plasma Research,
Bhat, Gandhinagar, India}
\email{kaw@ipr.res.in}

\begin{abstract}
Using the Berry-curvature modified kinetic equation we study instabilities in anisotropic chiral plasmas.
It is demonstrated that even for a very small value of the anisotropic parameter the chiral-imbalance 
instability is strongly modified. The instability is enhanced when the modes propagates in
the direction parallel to the anisotropy vector and 
it is strongly suppressed when the  modes propagate  in the perpendicular direction.
Further the instabilities in the jet-plasma system is also investigated. 
For the case when the modes are propagating in direction parallel to the stream  velocity we find that there
exist a new  branch of the dispersion relation arising due to the parity odd effects. 
We also show that the parity-odd interaction can enhance the streaming instability. 

\end{abstract}
\maketitle

\section{Introduction}

The scope of applying kinetic theory to understand variety of many-body problems arising in various branches of
physics is truly enormous \cite{Landau_kinetics}. The conventional Boltzmann or Vlasov equations
imply that the vector current associated with the gauge charges is conserved.
But till recently a very important class of physical phenomena associated with the 
CP-violation or the triangle-anomaly were left out of the purview of a kinetic theory.
In such a phenomenon the axial current is not conserved. 
It should be noted here that there exists a several models of hydrodynamics
which incorporates the effect of CP-violation \cite{Son:2009tf, Banerjee:2012iz, Jensen:2012jy, 
Kharzeev:2010gr}. But a hydrodynamical approach requires that
the system under consideration remains in a thermal and chemical equilibrium.  
However, many  applications of the chiral (CP-violating) physics
may involve a non-equilibrium situation e.g. during the early stages of
relativistic heavy-ion collisions. 
Therefore it is highly desirable to have a proper kinetic theory framework to tackle 
the CP-violating effect. 
Recently there has been a lot of progress in developing such a kinetic theory.
In Ref. \cite{Son:2012wh, Zahed:2012, Stephanov:2012, Chen:2013, Loganayagam:2012, Xiao:2005} 
it was shown that if the Berry curvature\cite{Berry} has nonzero flux across
the Fermi-surface then the particles on the surface can exhibit a chiral anomaly in presence of an external
electromagnetic field. In this formalism  chiral-current $j^\mu$ is not conserved and it can be attributed
to Adler-Bell-Jackiw anomaly \cite{adler69, bell69, Nielsen:1983rb}. It can be shown that
if a system of charged fermions does not conserve parity, it can develop an equilibrium electric current along an
applied external magnetic field \cite{vilenkin:80}. This is so called chiral-magnetic effect (CME). It has been
suggested that a strong magnetic field created in relativistic-heavy-ion experiments
can lead to CME in the quark-gluon plasma \cite{Fukushima:2008xe,Kharzeev08, Warringa08}. 
Indeed the recent experiments with STAR detector at Relativistic Heavy Ion Collider (RHIC)
qualitatively agree with a local parity violation. However,
more investigations are required to attribute this charge asymmetry with the CME \cite{Abelev:2009, Abelev:2010}.
The idea that a Berry-phase can influence the electronic properties [e.g. \cite{Xiao:2010} 
and references cited therein] is well-known in condensed matter literature and it can have applications in 
Weyl's semimetal \cite{kim13}, graphene \cite{Sasaki01} etc. 
There exists a deep connection between a CP-violating quantum field theory and 
the kinetic theory with the Berry curvature corrections. In Ref. \cite{Son:2012zy} it was shown
that the parity-odd and parity-even correlations calculated using the modified kinetic theory 
are identical with the perturbative results obtained in next-to-leading order hard dense loop approximation.

In this work we aim to apply the kinetic theory with the Berry curvature curvature
corrections to some non-equilibrium situations.
 We first note that the results obtained in Refs. \cite{Son:2012wh, Son:2012zy}
are limited to low temperature regime $T\ll \mu$, where $\mu$ is chiral chemical potential, when
the Fermi surface is well-defined. 
Recently Ref.\cite{Manuel:2013} argues that the domain of validity of the modified kinetic theory 
can be extended beyond the Fermi surface to include the effect of  finite temperature.   
As expected from the considerations
of quantum-field theoretic approach \cite{Itoyama:1983, Liu:1988, Nicola:1994} 
the parity-odd contribution remains temperature independent. 
Recently using the modified-kinetic theory \cite{Son:2012zy} in presence of the chiral imbalance
the collective modes in electromagnetic or quark-gluon plasmas were analyzed \cite{Akamatsu:2013}.
In such a system CP-violating effect can split transverse waves into two branches \cite{Nieves:1989}.
It was found in Ref. \cite{Akamatsu:2013} that in the quasi-static limit i.e. for $\omega \ll k$,
where $\omega$ and $k$ respectively denote frequency and wave-number of the transverse wave,
there exists an unstable mode. The instability can lead to the growth of Chern-Simons number
(or magnetic-helicity in plasma physics parlance) at expense of the chiral imbalance. 
Similar kind of instabilities were found in Refs. 
\cite{Redlich:1985, Rubakov:1985,Joyce:1997,Liane:2005, Boyarsky:2012} 
in different context. 

 In the present work we study collective modes
in anisotropic chiral plasmas. In many realistic situations in condensed matter
physics (see for example \cite{Tsitsadze:2009, Hu:1991}) and in plasma physics 
\cite{Krall} it is important to consider initial distribution function $n^0_{\bf p}$
to be anisotropic in the momentum space. 
It is well-known that momentum anisotropy can lead to
so called  Weibel instability of transverse waves in plasma which can generate
a magnetic field in the plasma \cite{Weibel:1959, Fried:1959}. The Weibel instability is closely related with
the streaming instabilities in plasma. Such instabilities
may play an important role in thermalization of  
the quark-gluon plasma created in relativistic heavy-ion collision experiments
\cite{Mro:1988,Bhatt:1994,Arnold:2003, Romatschke:2002, romatschke}. In this work we generalize
the modified kinetic theory to consider anisotropic chiral plasma.
In particular we consider two important cases: (i) when the 
distribution function $n^0_{\bf p}$ has a preferred direction (an anisotropy) in the momentum space.
(ii) when a stream of charged particles travel in a thermally equilibrated chiral-plasma.  
We believe that the results presented here will be useful in studying Weyl metals and quark-gluon plasmas
created in relativistic heavy-ion collisions.
The paper is organized as follows: In section II we give a brief introduction of the basic equations 
of the Berry-curvature modified kinetic and the Linear response theories. Section III
contains the case of Weibel-instability in anisotropic chiral-plasma. Section IV deals with
instability related with the jet-plasma interaction. Section VI 
contains summary and conclusions.

\section{Basic Equations}
 The Berry curvature modified collisionless kinetic (Vlasov) equation for distribution function 
$n_{\bf p}$ \cite{Son:2012zy} can be written as:
\beq
\label{eq:kinetic}
\dot n_{\bf p} + \frac{1}{1 + e {\bf B} \cdot {\bf \Omega}_{\bf p}}
\left[\left(e\tilde {\bf E} +e \tilde {\bf v} \times {\bf B}
+ (e^2 \tilde {\bf E} \cdot {\bf B}) {\bm \Omega}_{\bf p} \right)
\cdot \frac{\d n_{\bf p}}{\d {\bf p}}
+ \left(\tilde {\bf v} + e \tilde {\bf E} \times {\bm \Omega}_{\bf p}
+e (\tilde {\bf v} \cdot {\bm \Omega}_{\bf p}) {\bf B}
\right) \cdot \frac{\d n_{\bf p}}{\d {\bf x}}
\right]=0,
\nonumber \\
\eeq
where $\tilde {\bf v} = \d{\epsilon_{\bf p}}/\d {\bf p}$,
$e{\tilde {\bf E}}=e{\bf E} - \d{\epsilon_{\bf p}}/\d {\bf x}$, $\mathbf{\epsilon_{p}}=p(1-e{\mathbf{B\cdot\Omega_{p}}})$ and 
${\mathbf{\Omega_{p}}}=\pm{\mathbf{p}}/{2 p^{3}}$. Here $\pm$ sign corresponds to right and lefted handed 
fermions respectively. In absence of the Berry curvature term (i.e. $\Omega_{\bf p}$=0) $\epsilon_{\bf{p}}$ is independent of x,   
Eq.(\ref{eq:kinetic}) reduces to the standard Vlasov equation. Particle density $n$ can be defined as 
 \beq
\label{eq:n}
n = \int \frac{d^3p}{(2\pi)^3}(1+e{\bf B}\cdot{\bm \Omega}_{\bf p})n_{\bf p},
\eeq
\noindent whereas the current density $\bf j$ can be defined as:

\beq
\label{eq:j}
{\bf j} = -\int \frac{d^3p}{(2\pi)^3}
\left[\epsilon_{\bf p}{\mathbf{\partial}}_{\mathbf{p}}n_{\mathbf{p}}
+ e\left({\bm \Omega}_{\bf p} \cdot {\mathbf{\partial}}_{\mathbf{p}}n_{\mathbf{p}}\right) \epsilon_{\bf p} {\bf B}
+ \epsilon_{\bf p} {\bm \Omega}_{\bf p} \times {\mathbf{\partial}}_{\mathbf{x}}n_{\mathbf{p}} \right] 
+e {\bf E} \times {\bm \sigma},
\eeq

\noindent where ${\mathbf{\partial}}_{\mathbf{P}}=\frac{\d}{\d {\bf p}}$ and ${\mathbf{\partial}}_{\mathbf{x}}=\frac{\d}{\d {\bf x}}$.
The last term on the right hand side of the above equation represents the anomalous Hall current with $\sigma$ given as follows: 
\beq
\label{eq:sigma}
{\bm \sigma} = \int \frac{d^3p}{(2\pi)^3} {\bm \Omega}_{\bf p} n_{\bf p}.
\eeq\

\subsection{Maxwell Equation, Propagator and Dispersion relation}
Using the above expression for the number and current densities one can write the 
Maxwell equation as,
\begin{equation}
\partial_{\nu}F^{\nu\mu}=j^{\mu}_{ind}+j^{\mu}_{ext}. 
\label{Maxwell:1}
\end{equation}
 Here $j^{\mu}_{ext}$ is an external current.  The induced current $j^{\mu}_{ind}$ can be expressed in terms 
 of gauge field $A_{\nu}(k)$ via linear response theory in Fourier space as,
\begin{equation}
j^{\mu}_{ind}=\Pi^{\mu\nu}(K)A_{\nu}(K),
\label{linresponse}
\end{equation}
\noindent
where $\Pi^{\mu\nu}(K)$ is the retarded self energy in Fourier space.
Here we have denoted a Fourier transform any quantity $F(x,t)$ by
$ F(K)=\int{d^4x} e^{-i(\omega{t}-\mathbf{k\cdot{x}})}F(x,t)$.  Now one can write Eq.( \ref{Maxwell:1})
in the Fourier space as
\begin{equation}
[K^{2}g^{\mu\nu}-K^{\mu}K^{\nu}+\Pi^{\mu\nu}(K)]=-j^{i}_{ext}(K). 
\end{equation}
\noindent
By choosing temporal gauge $A_{0}=0$ we can write the above equation as,
\begin{equation}
 [\Delta^{-1}(K)]^{ij}E^{j}=[({k^2}-{\omega}^{2})\delta^{ij}-k^{i}k^{j}+\Pi^{ij}(K)]E^{j}=i\omega{j^{i}_{ext}}(k). 
\label{eq1}
\end{equation}
\noindent
From this one can define
\beq
[\Delta^{-1}(K)]^{ij}=({k^2}-{\omega}^{2})\delta^{ij}-k^{i}k^{j}+\Pi^{ij}(K).
\label{invprop}
\eeq
By finding inverse of the object $[\Delta^{-1}(K)]^{ij}$ one can obtain the expression for the propagator
$[\Delta(K)]^{ij}$ whose poles can give the dispersion relation.
In a linear response theory we are interested in the induced current by a linear-order deviation
in the gauge field. We follow  the power counting scheme of Ref. \cite{Son:2012zy}: gauge field $A_\mu=O(\epsilon)$
and derivatives $O(\delta)$, where $\epsilon$  and $\delta$ are small and independent parameters.  In this scheme
one considers deviations in the current and the distribution function up to $O(\epsilon\delta)$. Under this counting scheme
one can write the kinetic equation as:
\begin{equation}
(\partial_{t}+{\mathbf{v}}\cdot{\mathbf{\partial_{x}}})n_{\mathbf{p}}+(e{\mathbf{E}}+e{\mathbf{v}}\times{\mathbf{B}}-
{{\mathbf{\partial}}_{\mathbf{x}}}\mathbf{\epsilon_{p}})\cdot{\mathbf{\partial}}_{\mathbf{p}}n_{\mathbf{p}}=0
\label{kineqwbcc}
\end{equation}
where $\mathbf{v}=\frac{\bf{p}}{p}$.   

\section{Collective Modes in Anisotropic Chiral Plasma}

 We consider the equilibrium distribution of the form $n^{0}_{\mathbf{p}}=1/[e^{(\mathbf{\epsilon_{p}}-\mu)/T}+1]$. 
Following the power counting scheme that we have introduced above one can write:
\beq 
n^{0}_{\mathbf{p}}=n^{0(0)}_{\mathbf{p}}+e n^{0(\epsilon\delta)}_{\mathbf{p}},
\label{equdistr1}
\eeq
where, 
$n^{0(0)}_{\mathbf{p}}=\frac{1}{[e^{(p-\mu)/T}+1]}$
and $n^{0(\epsilon\delta)}_{\mathbf{p}}=\left(\frac{{\mathbf{B}\cdot\mathbf{v}}}{2 p T}\right)
\frac{e^{(p-\mu)/T}}{[e^{(p-\mu)/T}+1]^2}$. 
In order to bring in effect of anisotropy we follow the arguments of Ref. \cite{romatschke}. It is assumed that
the anisotropic equilibrium distribution function can be obtained from a spherically symmetric distribution function
by rescaling of one direction in the momentum space. Thus we assume that there is a momentum anisotropy 
in direction of a unit vector ${\mathbf{\hat{n}}}$. Noting that $p=|{\mathbf p}|$, we replace 
$p\rightarrow \sqrt{{\mathbf p}^{2}+\xi({\mathbf p}\cdot{\mathbf{\hat{n}}})^2}$ in Eq.(\ref{equdistr1}) to get 
the anisotropic distribution function. 
Here $\xi$ is an adjustable anisotropy parameter satisfying a condition $\xi>-1$.
It is convenient to define
a new variable  $\tilde{p}$ such that $\tilde{p}=p\sqrt{1+\xi({\mathbf v}\cdot{\mathbf{\hat{n}}})^2}$. Using this 
new variable one can write $n^{0(0)}_{\mathbf{p}}=\frac{1}{[e^{(\tilde{p}-\mu)/T}+1]}$ and  
$n^{0(\epsilon\delta)}_{\mathbf{p}}=\left(\frac{{\mathbf{B}\cdot\mathbf{v}}}{2 \tilde{p} T}\right)
\frac{e^{(\tilde{p}-\mu)/T}}{[e^{(\tilde{p}-\mu)/T}+1]^2}$. 

 The anomalous Hall current term in Eq.(\ref{eq:j}) can vanish 
if the distribution function is spherically symmetric in the momentum space. However, for an anisotropic
distribution function this may not be true in general. Since the Hall-current term depends on  electric field, 
it can be of order $O(\epsilon\delta)$ or higher.  
As we are interested in finding  deviations in current and distribution function up to order $O (\epsilon\delta)$,
only $n^{0(0)}_{\mathbf{p}}$ would contribute to the Hall current term. 
Next, we consider ${\bm \sigma}$ from Eq.(\ref{eq:sigma}) which can be written as
\beq
{\bm \sigma}=\frac{1}{2}\int d\Omega d\tilde{p}\frac{\mathbf v}{[1+\xi(\mathbf{v}\cdot\mathbf{\hat{n}})]^{1/2}}
\frac{1}{(1+e^{(\tilde{p}-\mu)/T})}.\label{sigma1}
\eeq    
\noindent
Since $\mathbf{v}$ is a unit vector one can express $\mathbf{v}=(sin\theta cos\phi, sin\theta sin\phi, cos\theta)$
in spherical coordinates. By choosing $\mathbf{\hat{n}}$ in $z-$direction, without any loss of generality,
one can have $\mathbf{v}\cdot\mathbf{\hat{n}}=cos\theta$. Thus the angular integral in the above equation
becomes $\int d(cos\theta)d\phi\frac{\mathbf{v}}{(1+\xi cos^2\theta)^{1/2}}$. Therefore 
$\sigma_{x}$ and $\sigma_{y}$ components of Eq.(\ref{sigma1}) will vanish as $\int^{2\pi}_0 sin\phi d\phi=0$ and
$\int^{2\pi}_0 cos\phi d\phi=0$. While $\sigma_{z}$ will vanish because integration with respect to $\cos\theta$ variable 
will yield it ($\sigma_{z}$) to be zero. 
Thus the anomalous Hall current term will not contribute for the problem at the hand.   
Now one can write  current ${\mathbf j}$ as follows, 
\begin{equation}
\mathbf{j}=-e\int \frac{d^{3}p}{(2\pi)^3}[\mathbf{\epsilon_{p}}{\mathbf{\partial}}_{\mathbf{p}}n_{\mathbf{p}}+
e({\mathbf{\Omega_{p}}}\cdot{\mathbf{\partial}}_{\mathbf{p}}n_{\mathbf{p}}){\mathbf{\epsilon_{p}}}{\mathbf{B}}+
\mathbf{\epsilon_{p}}{\mathbf{\Omega_{p}}}\times{\mathbf{\partial}}_{\mathbf{x}}n_{\mathbf{p}}].
\label{curntwithbcc}
\end{equation} 
The distribution function can be decomposed into separate scales as follows,
\begin{equation}
n_{\mathbf{p}}=n^{0}_{\mathbf{p}}+e (n^{(\epsilon)}_{\mathbf{p}}+n^{(\epsilon\delta)}_{\mathbf{p}}). 
\label{expansion of distribution function in terms of `epsilon' and `epsilondelta'}
\end{equation}
Now the kinetic equation (\ref{kineqwbcc}) can be split into two equations valid    
at  $O(\epsilon)$ and  $O(\epsilon\delta)$ scales as written below,
\begin{equation}
(\partial_{t}+{\mathbf{v}}\cdot{\mathbf{\partial_{x}}})n^{(\epsilon)}_{\mathbf{p}}=-({\mathbf{E}}+
{\mathbf{v}}\times{\mathbf{B}})\cdot{\mathbf{\partial}}_{\mathbf{p}}n^{0(0)}_{\mathbf{p}}
\label{kEqnfrdistbtnfnaodrepsilon}
\end{equation}
\begin{equation}
(\partial_{t}+{\mathbf{v}}\cdot{\mathbf{\partial_{x}}})(n^{0(\epsilon\delta)}_{\mathbf{p}}+n^{(\epsilon\delta)}_{\mathbf{p}})=
-\frac{1}{e}{{{\mathbf{\partial}}_{\mathbf{x}}}\mathbf{\epsilon_{p}}}\cdot{\mathbf{\partial}}_{\mathbf{p}}{n^{0(0)}_{\mathbf{p}}}
\label{kEqnfrdistbtnfnaodrepsilondelta}
\end{equation}
Equation for the current defined in Eq.(\ref{curntwithbcc}) 
can also split into  $O(\epsilon)$ and $O(\epsilon\delta)$ scales  as given below,
\begin{equation}
 \mathbf{j^{\mu(\epsilon)}}=e^2\int \frac{d^{3}p}{(2\pi)^3}v^{\mu}n^{(\epsilon)}_{\mathbf{p}}
\label{curntwithbccioepsilon}
\end{equation}
\begin{equation}
 \mathbf{j^{i(\epsilon\delta)}}=e^2\int \frac{d^{3}p}{(2\pi)^3}\left[v^{i}n^{(\epsilon\delta)}_{\mathbf{p}}-
 \left(\frac{{v^{j}}}{2 p}\frac{\partial{n^{0(0)}_{\mathbf{p}}}}{\partial{p^{j}}}\right){{B^{i}}}
 -\epsilon^{ijk}\frac{v^{j}}{2 p}
  \frac{\partial{n^{(\epsilon)}_{\mathbf{p}}}}{\partial{x^{k}}}\right]
  \label{curntwithbccioepsilondelta}
\end{equation}
The self-energy or polarization tensor in Eq.(\ref{linresponse}) contains parity-even $\Pi^{ij}_{+}$ and 
parity-odd $\Pi^{ij}_{-}$
parts and thus one can write $\Pi^{ij}=\Pi^{ij}_{+}+\Pi^{ij}_{-}$.  
Using Eqs. (\ref{linresponse}, \ref{kEqnfrdistbtnfnaodrepsilon}, \ref{kEqnfrdistbtnfnaodrepsilondelta}, 
\ref{curntwithbccioepsilon}, \ref{curntwithbccioepsilondelta}) one can obtain the expression for
$\Pi^{ij}_{+}$ and $\Pi^{ij}_{-}$ as:  
\begin{equation}
 {\Pi^{ij}_{+}(K)}=m^{2}_{D}\int \frac{d\Omega}{4 \pi}
 \frac{{v^{i}}({v^{l}}+\xi(\mathbf{v\cdot{\hat{n}}})\hat{n}^l)}{(1+\xi(\mathbf{v\cdot{\hat{n}}})^2)^2}
 \left(\delta^{jl}+\frac{v^{j}k^{l}}{v.k+i\epsilon}\right),
\label{selfenergyoepsilon}
\end{equation}
\begin{align}
 {\Pi^{im}_{-}(K)}=C_{E}\int 
 \frac{d\Omega}{4 \pi}\Bigg[\frac{i{\epsilon}^{jlm}k^{l}v^{j}v^{i}(\omega+
 \xi(\bf{v\cdot\hat{n}})(\bf{k.\hat{n}}))}{(v.k+i{\epsilon})(1+\xi(\mathbf{v\cdot{\hat{n}}})^2)^{3/2}}+
 \left(\frac{{v^{j}}+\xi(\mathbf{v\cdot{\hat{n}}})\hat{n}^j}
 {(1+\xi(\mathbf{v\cdot{\hat{n}}})^2)^{3/2}}\right){i{\epsilon}^{iml}k^{l}v^{j}}\nonumber\\-{i{\epsilon}^{ijl}k^{l}v^{j}}
 \left(\delta^{mn}+\frac{v^{m}k^{n}}{v.k+i\epsilon}\right)
 \left(\frac{{v^{n}}+\xi(\mathbf{v\cdot{\hat{n}}})\hat{n}^n}
 {(1+\xi(\mathbf{v\cdot{\hat{n}}})^2)^{3/2}}\right)
 \Bigg] 
\label{selfenergyoepsilondelta}
\end{align}
where,
\begin{eqnarray}
 m^{2}_{D}=-\frac{e^2}{2\pi^{2}}\int_{0}^{\infty}d\tilde{p}{\tilde{p}}^{2}
 \left[\frac{\partial{n^{0(0)}_{\mathbf{\tilde{p}}}}(\tilde{p}-\mu)}{\partial{\tilde{p}}}+
 \frac{\partial{n^{0(0)}_{\mathbf{\tilde{p}}}}(\tilde{p}+\mu)}{\partial{\tilde{p}}}\right]\nonumber\\
 C_{E}=-\frac{e^2}{4\pi^{2}}\int_{0}^{\infty}d\tilde{p}{\tilde{p}}
 \left[\frac{\partial{n^{0(0)}_{\mathbf{\tilde{p}}}}(\tilde{p}-\mu)}{\partial{\tilde{p}}}-
 \frac{\partial{n^{0(0)}_{\mathbf{\tilde{p}}}}(\tilde{p}+\mu)}{\partial{\tilde{p}}}\right]
\label{thermalmass}.
\end{eqnarray}
\noindent
We would like to mention that one can 
write the total current $\bf{j}=\bf{j}^{\epsilon}+\bf{j}^{\epsilon\delta}$ where $\bf{j}^{\epsilon}$ 
and $\bf{j}^{\epsilon\delta}$ respectively denote the 
vector and axial currents. $\bf{j}^{\epsilon}$ gives contribution of order of the square of plasma frequency or $m^2_{D}$. The plasma 
frequency contains additive contribution from the densities of 
all species i.e. right-handed particle/antiparticles and left-handed particles/antiparticles. The axial current 
arises due to chiral imbalance its contribution from each plasma specie, depends upon $e\vec{\Omega_{p}}$. 
Since $e\vec{\Omega_{p}}$ can change sign depending on the plasma specie therefore definition of 
$C_{E}$ contains both positive and negative signs. Consequently a relative signs of fermion and anti-fermion are different 
in $m^2_{D}$ and $C_{E}$. After performing above integrations one can 
get $m^{2}_{D}=e^2\left(\frac{\mu^2}{2\pi^2}+\frac{T^2}{6}\right)$ and $C_E=\frac{e^2 \mu}{4\pi^2}$. Where $\mu$ is 
the chemical potential for chiral fermions. It is to be noted that $C_{E}=0$ when there is no chiral-chemical 
potential where as $m^2_{D}\neq0$.
It can also be noticed that the terms with anisotropy parameter $\xi$ are contributing in parity-odd part
of the self-energy. Further we would like to note that we have used expression for right-handed current by
adding particle and anti-particle contributions in obtaining Eqs.(\ref{selfenergyoepsilon}-\ref{thermalmass}).
Contributions from the left-handed particles can be just be added very easily and  Eq.(\ref{thermalmass}) will
exactly match with the one given in Ref.\cite{Manuel:2013}. Introduction of chemical potential $\mu$ for chiral fermions requires 
some qualification. Physically the chiral 
chemical potential imply an imbalance between the right handed and left handed fermion. This in turn related to 
the topological charge\cite{Fukushima:2008xe, Redlich:1985}. It should be noted here that due to the axial anomaly 
chiral chemical potential is not associated with any conserved charge. It can still be regarded as `chemical potential' 
if its variation is sufficiently slow\cite{Akamatsu:2013}.

\subsection{Finding the Poles of $[\Delta(K)]^{ij}$ or Dispersion relation}

In order to get the expression for the propagator $\Delta^{ij}$ it is necessary to
write $\Pi^{ij}$ in a tensor decomposition. For the present problem we need six independent 
projectors. For an isotropic parity-even plasmas one may need the transverse $P^{ij}_{T}$
and the longitudinal $P^{ij}_{L}$ tensor projectors. Due to anisotropy coming due to the presence
direction ${\mathbf n}$ one needs two more projectors $P^{ij}_{n}$ and $P^{ij}_{kn}$ \cite{Kobes:1991}.
To account for parity odd effect we have to include two anti-symmetric operators $P^{ij}_{A}$
and $P^{ij}_{An}$. Thus we write $\Pi^{ij}$ into the basis spanned by the above six operators as:
\begin{equation}
 \Pi^{ij}=\alpha{P}^{ij}_{T}+\beta{P}^{ij}_{L}+\gamma{P}^{ij}_{n}+\delta{P}^{ij}_{kn}+\lambda{P}^{ij}_{A}+\chi{P}^{ij}_{An}.
 \label{seexpan}
\end{equation}
where, ${P}^{ij}_{T}=\delta^{ij}-k^{i}k^{j}/{k^{2}}$, ${P}^{ij}_{L}=k^{i}k^{j}/{k^{2}}$, 
${P}^{ij}_{n}={\tilde{n}}^i{\tilde{n}}^j/{\tilde{n}}^2$, 
${P}^{ij}_{kn}=k^{i}{\tilde{n}}^{j}+k^{j}{\tilde{n}}^{i}$, 
${P}^{ij}_{A}=i\epsilon^{ijk}{\hat{k}}^k$ and 
${P}^{ij}_{An}=i\epsilon^{ijk}{\tilde{n}}^k$. $\alpha$,$\beta$, $\gamma$, $\delta$ $\lambda$ and $\chi$ are 
some scalar functions of $k$ and $\omega$ which are yet to be determined.

Similarly  we can write $[\Delta^{-1}(k)]^{ij}$ appearing in Eq.(\ref{eq1}) as
\begin{equation}
 [\Delta^{-1}(K)]^{ij}=C_{T}{P}^{ij}_{T}+C_{L}{P}^{ij}_{L}+C_{n}{P}^{ij}_{n}+C_{kn}{P}^{ij}_{kn}+C_{A}{P}^{ij}_{A}+C_{An}{P}^{ij}_{An}.
 \label{invpropexpan}
\end{equation}
Using Eqs.(\ref{invprop}, \ref{seexpan}, \ref{invpropexpan}) one can express relationship between $C$'s and the scalar functions
defined in Eq.(\ref{seexpan}) as:
\begin{eqnarray}
 C_{T}=k^{2}-{\omega}^{2}+\alpha,
 C_{L}=-{\omega}^{2}+\beta,
 C_{n}=\gamma,
 C_{kn}=\delta,
 C_{A}=\lambda,
 C_{An}=\chi.\label{eq2}
\end{eqnarray}
It should be noted that
 $\alpha=({P}^{ij}_{T}-{P}^{ij}_{n})\Pi^{ij}$, $\beta={P}^{ij}_{L}\Pi^{ij}$, 
$\gamma=(2{P}^{ij}_{n}-{P}^{ij}_{T})\Pi^{ij}$, $\delta=\frac{1}{2 k^{2}{\tilde{n}}^2}{P}^{ij}_{kn}\Pi^{ij}$
$\lambda=-\frac{1}{2}{P}^{ij}_{A}\Pi^{ij}$ and $\chi=-\frac{1}{2{\tilde{n}}^2}{P}^{ij}_{An}\Pi^{ij}$. 
In the limit $\xi\rightarrow0$, using Eqs.(\ref{selfenergyoepsilon}-\ref{selfenergyoepsilondelta}) one calculate
$\alpha_{\arrowvert_{\xi=0}}=\Pi_{T}$, 
$\beta_{\arrowvert_{\xi=0}}=\frac{\omega^2}{k^2}\Pi_{L}$, $\gamma_{\arrowvert_{\xi=0}}=0$, $\delta_{\arrowvert_{\xi=0}}=0$, 
$\lambda_{\arrowvert_{\xi=0}}=-\frac{\Pi_{A}}{2}$ and $\chi_{\arrowvert_{\xi=0}}=0$ where,
\begin{eqnarray}
\Pi_{T}=m^{2}_{D}\frac{\omega^{2}}{2 k^{2}}\left[1+\frac{k^{2}-{\omega}^{2}}{2\omega k}\ln\frac{\omega+k}{\omega-k}\right],\nonumber\\
\Pi_{L}=m^{2}_{D}\left[\frac{\omega}{2 k}\ln\frac{\omega+k}{\omega-k}-1\right],\nonumber\\
\Pi_{A}=-2 k C_{E}\left(1-\frac{{\omega}^2}{k^2}\right)\left[1-\frac{\omega}{2 k}\ln\frac{\omega+k}{\omega-k}\right].
\label{isotropic:eq}
\end{eqnarray}
Scalar functions $\Pi_T$, $\Pi_{L}$ and $\Pi_A$ respectively representing the transverse, longitudinal
and the axial part of the self-energy decomposition in the tensorial basis in the $\xi=0$ limit as defined in Ref. \cite{Akamatsu:2013}. 

Next, we expand $[\Delta(K)]^{ij}$ in the tensor projector basis as:
\begin{equation}
[\Delta(K)]^{ij}=a{P}^{ij}_{L}+b{P}^{ij}_{T}+c{P}^{ij}_{n}+d{P}^{ij}_{kn}+e{P}^{ij}_{A}+f{P}^{ij}_{An}.
\end{equation}
It is rather easy but rather cumbersome to express 
the coefficients $a$, $b$, $c$, $d$, $ e$ and $f$ in terms of the coefficients $C$'s appearing in Eq.(\ref{invpropexpan}) using the relation 
$[\Delta^{-1}(K)]^{ij} [\Delta(K)]^{jl}=\delta^{il}$.
The dispersion relation can be obtained by equating denominators of the expressions 
for $a$, $b$, $c$, $d$, $ e$ and $f$ with zero. In the present case denominator 
for $a$, $b$, $c$, $d$, $ e$ and $f$ is same therefore the dispersion relation can be written as:
\begin{equation}
{2 k {\tilde{n}}^{2}C_{A}C_{An}C_{kn}+C^{2}_{A}C_{L}+{\tilde{n}}^{2}C^{2}_{An}(C_{n}
+C_{T})-C_{T}(-k^{2}{\tilde{n}}^{2}C^{2}_{kn}+C_{L}(C_{n}+C_{T}))}=0. 
\label{dispersionrelation}
\end{equation}
 This general form of the dispersion relation is quite complicated. The expression for $\alpha$,
$\beta$, $\gamma$ and $\delta$ exactly match with those given in Ref. \cite{romatschke}. 
The new contribution comes in terms of the coefficients $\lambda$ and $\chi$ which contains the
effect of parity violation. But note that the standard criteria for the plasma instability 
(Weibel) \cite{Arnold:2003} are not applicable here because of the parity violation. 
For small anisotropy parameter $\xi$ it is possible to evaluate all the
integrals in the dispersion relation analytically.

\subsection{Analysis of the collective modes in small $\xi$ limit}

Using ${\bf{\hat{k}}.{\hat{n}}}=cos{\theta_n}$ we can express $\alpha$, $\beta$, $\gamma$, $\delta$, $\lambda$ and $\chi$
up to linear order in $\xi$ as follows,
\begin{eqnarray}
\alpha&=&\Pi_{T}+\xi\left[\frac{z^2}{12}(3+5\cos{2\theta_{n}})m^{2}_{D}-\frac{1}{6}(1+\cos{2\theta_{n}})m^{2}_{D}+
\frac{1}{4}\Pi_{T}\left((1+3\cos{2\theta_{n}})-z^{2}(3+5\cos2\theta_{n})\right)\right];\nonumber\\
z^{-2}\beta&=&\Pi_{L}+\xi\left[\frac{1}{6}(1+3\cos{2\theta_{n}})m^{2}_{D}+\Pi_{L}
\left(\cos{2\theta_{n}}-\frac{z^2}{2}(1+3\cos{2\theta_{n}})\right)\right];\nonumber\\
\gamma&=&\frac{\xi}{3}(3\Pi_{T}-m^{2}_{D})(z^2-1)\sin^2{\theta_{n}};\nonumber\\
\delta&=&\frac{\xi}{3 k}(4 z^{2} m^{2}_{D}+3\Pi_{T}(1-4z^{2}))\cos{\theta_{n}};\nonumber\\
\lambda&=&-\frac{\mu k e^2}{4\pi^2}\left[(1-z^2)\frac{\Pi_{L}}{m^{2}_{D}}\right]-\xi\frac{\mu k e^2}{32 \pi^2}
\Bigg[(1-z^2)\frac{\Pi_{L}}{m^{2}_{D}}\left((1+7\cos{2\theta_{n}})-3z^2(1+3\cos{2\theta_{n}})\right)\nonumber\\&+&
\frac{1}{3}(1+11\cos{2\theta_{n}})-z^2(3+5\cos{2\theta_{n}})
\Bigg];\nonumber\\
\chi&=&\xi\left[f(\omega,k)\right],
\label{eq3}
\end{eqnarray}
\noindent
where $z=\frac{\omega}{k}$ and $f(\omega,k)$ is some function $k$ and $\omega$. But in the present analysis 
its exact form of $f(\omega,k)$ form may not be required.
Using the above equations and Eqs. (\ref{isotropic:eq}, \ref{eq2}) one can finally express
Eq.(\ref{dispersionrelation})  in terms of $k$ and $\omega$. 
One can notice from Eq.(\ref{eq3}) that the most significant contribution for  
$\gamma$, $\delta$, $\lambda$ and $\chi$ is $O(\xi)$. Thus in the present scheme of 
approximation one can write Eq.(\ref{dispersionrelation}) up to $O(\xi)$ as: 
\begin{equation}
 C^{2}_{A}C_{L}-C_{T}C_{L}(C_{n}+C_{T}))=0, \label{dispersionrelation1}
\end{equation}
which in turn can give following two branches of the dispersion relation,
\begin{eqnarray}
\label{dispersionrelation2}
C^{2}_{A}-C^{2}_{T}-C_{n}C_{T}=0,\\
C_{L}=0.\label{dispersionrelation3}
\end{eqnarray}
First, we would like to note that when $C_{A}=0$, Eqs.(\ref{dispersionrelation2}-\ref{dispersionrelation3}) reduces to exactly the same
dispersion relation discussed in Ref.\cite{romatschke} for an anisotropic plasma where
there is no parity violating effect.
Let us consider Eq.(\ref{dispersionrelation2}), it can be written as:
\begin{equation}
(k^2-\omega^2)^2+(k^2-\omega^2)(2\alpha+\gamma)+\alpha^2+\alpha\gamma-\lambda^2=0.
\end{equation}
This equation is a quadratic equation in $(k^2-\omega^2)$ and it's solutions can be written as,
\begin{equation}
(k^2-\omega^2)=\frac{-(2\alpha+\gamma)\pm2\lambda}{2}.\label{eq4}
\end{equation}
It is of particular interest to consider 
the quasi-static limit $\left|{\omega}\right|<<k$, in this limit expressions for $\alpha\sim\Pi_T$ and 
$\beta\sim\frac{\omega^2}{k^2}\Pi_L$ and $\lambda\sim-\frac{\Pi_A}{2}$. Now $\Pi_L$, $\Pi_T$ and $\Pi_A$ can be obtained 
by expanding Eq.(\ref{isotropic:eq}) in the quasi static limit as:
\begin{eqnarray}
{\Pi_{T}}_{\arrowvert_{\left|{\omega}\right|<<k}}=\left(\mp{i}\frac{\pi}{4}\frac{\omega}{k}\right)m^{2}_{D};\nonumber\\
{\Pi_{L}}_{\arrowvert_{\left|{\omega}\right|<<k}}= m^{2}_{D}\left[\mp{i}\frac{\pi}{2}\frac{\omega}{k}-1\right]\nonumber\\
{\Pi_{A}}_{\arrowvert_{\left|{\omega}\right|<<k}}=
-\frac{\mu k e^2}{2 \pi^2}\left(\frac{{\Pi_{L}}_{\arrowvert_{\left|{\omega}\right|<<k}}}{m^2_D}\right)\label{quasilimitofpit}
\end{eqnarray}
Therefore in quasi-stationary limit one can write positive branch of Eq.(\ref{eq4}) as 
$\omega={i}\rho(k)$ where, $\rho(k)$ is given by the following expression,
\begin{equation}
\rho(k)=\pm\frac{\left(\frac{4\alpha_{e}^3\mu^3}{\pi^{4}m^2_{D}}\right)k^2_{N}\left[1-k_{N}+
\frac{\xi}{12}\left(1+5\cos{2\theta_{n}}\right)
+\frac{\xi}{12}\left(1+3\cos{2\theta_{n}}\right)\frac{\pi^2 m^{2}_{D}}{\mu^2\alpha_{e}^2 k_N}\right]}
{{\left[1+\frac{2\mu^2 \alpha_{e}^2 k_N}{\pi^2 m^2_{D}}(1-\frac{\xi}{4})+\xi\cos{2\theta_{n}}\left(1-
\frac{7 \mu^2 \alpha_{e}^2 k_N}{2 \pi^{2} m^2_{D}}\right)\right]}}
\label{eq5} 
\end{equation}
Thus $\omega$ is purely an imaginary number and its real-part is zero i.e. $Re(\omega)=0$. Eq. (\ref{eq5}) have positive 
and negative signs. The negative sign  Eq. (\ref{eq5}) is unphysical. This can be seen from the fact 
when $\Pi_{A}=0$ and $\xi=0$ the negative 
branch of Eq. (\ref{eq5}) gives an instability. But in this case there is no source of free energy either in terms 
chiral imbalance or in terms of anisotropy in momentum space. Henceforth we ignore the negative sign.
Further we have defined $\alpha_{e}=\frac{e^2}{4\pi}$ as the electromagnetic coupling and $k_{N}=\frac{\pi}{\mu\alpha_{e}}k$ normalized 
wave number. 
Positive  $\rho(k)>0$ implies an instability as $e^{-i(i\rho(k))t}\sim e^{+\rho(k)t}$. 

In the denominator of  Eq.(\ref{eq5}) the terms containing $\alpha$ can dropped as compared to unity for $k_{N}\sim O(1)$
because $\frac{\mu^2 \alpha_{e}^2 k_N}{\pi^2 m^{2}_{D}}<\frac{\alpha_{e}}{\pi^2}k_{N}\ll1$. The denominator now can be written as 
$(1+\xi \cos2\theta_{n})$. One can expand the denominator in powers of 
$\xi$ and keep only linear term in Eq.(\ref{eq5}). Next one can notice that among all $\xi$-dependent terms in the numerator the  
term with $\alpha_{e}^{-2}$ will dominate. Thus one can write, 

\begin{equation}
\rho(k)={\left(\frac{4\alpha_{e}^3\mu^3}{\pi^{4}m^2_{D}}\right)k^2_{N}\left[1-k_{N}+
\frac{\xi}{12}\left(1+3\cos{2\theta_{n}}\right)\frac{\pi^2 m^{2}_{D}}{\mu^2\alpha_{e}^2 k_N}\right]}.
\label{result:1}
\end{equation}
One can get an upper bound on $\xi$ by 
substituting $\arrowvert\omega\arrowvert=\rho(k)=k$ in above 
equation. For $\theta_{n}=0$ and $k_{N}=1$ upper bound is $\xi=\frac{3\pi}{4}$.
Before we analyze the interplay between the chiral-imbalance and the Weibel instabilities, it is instructive to qualitatively
understand their origin.  First consider the chiral-imbalance instability. For a such a plasma `chiral-charge' density $n$
is given by $\partial_t n+{\bf  \nabla \cdot j}=\frac{2\alpha} {\pi} {\bf E \cdot B} $.  From this one can estimate the axial charge density
$n\sim \alpha k A^2$ where $A$ is the gauge-field. The number and energy densities of the plasma respectively given by $\mu T^2$ and $\mu^2 T^2$.  The typical energy for the gauge field $\epsilon_A\sim k^2A^2$. From the above value of the wave-vector
it can be seen that $\epsilon_A=\mu^2 T^2\frac{T^2}{\alpha^2 A^2}$. Thus for $\frac{T^2}{\alpha^2} <A$, the energy in the
gauge field is lower than the energy of the particle. This leads to the chiral-imbalance instability\cite{Akamatsu:2013,Joyce:1997}.
The Weibel instability arises when the equilibrium distribution function of the plasma has anisotropy in the momentum 
space\cite{Weibel:1959, Fried:1959}.
The anisotropy  in the momentum space can be regarded as anisotropy in temperature. Suppose there is plasma which is hotter
in $y$-direction than $x$ or $z$ direction.  If in this situation a disturbance with a magnetic-field $B=B_0cos(k x)$ which 
arises  say from noise,  then the Lorentz-force can produce current-sheets where the magnetic field changes its sign.
The current-sheet in turn enhancing the original magnetic field \cite{Weibel:1959, Fried:1959}

The unity term in the square bracket of Eq.(\ref{result:1}) is due to the chiral imbalance, while $\xi$ 
dependent term is due to momentum anisotropy. First consider the case when $\xi=0$, the above equation reduces 
to the dispersion-relation 
in Ref.\cite{Akamatsu:2013} describing the instability due to the chiral imbalance in the range $0<k_N<1$. Next, 
we consider the case when there is no chiral imbalance, in this case Eq.(\ref{result:1}) can be written as,
\begin{equation}
\rho(k)={\left(\frac{4\alpha_{e}^3\mu^3}{\pi^{4}m^2_{D}}\right)k^2_{N}\left[-k_{N}+
\frac{\xi}{12}\left(1+3\cos{2\theta_{n}}\right)\frac{\pi^2 m^{2}_{D}}{\mu^2\alpha_{e}^2 k_N}\right]}.
\end{equation}
This gives unstable modes for Weibel instability, in the quasi-static limit $\arrowvert \omega \arrowvert<<k$, when the 
following condition on $k_{N}$ is satisfied,
\begin{equation}
 0<k_N<\sqrt{\frac{\xi}{12}\left(1+3 cos2\theta_n\right)}\left(\frac{\pi m_D}{\mu\alpha_{e}}\right).
\label{condition3}
\end{equation}
Thus the chiral imbalance and Weibel instabilities have overlapping ranges. Maximum growth rates for  
the chiral instability is $\Gamma_{ch}=\frac{4}{27}\left(\frac{4\alpha_{e}^3\mu^3}{\pi^{4}m^2_{D}}\right)$ and  
for Weibel instability 
$\Gamma_{w}=2\left(\frac{4\alpha_{e}^3\mu^3}{\pi^{4}m^2_{D}}\right)\left(\frac{\xi\pi^2 m^{2}_{D}}{9\mu^2\alpha_{e}^2}\right)^{3/2}$ 
at $\theta_{n}=0$. 
For $\mu\sim T$, 
$m^{2}_D/\mu^2=\frac{2}{3}\frac{\alpha_{e}}{\pi}(3+\pi^2)$ and the ratio 
$\Gamma_{w}/\Gamma_{ch}\sim\frac{1}{2}\left[\frac{2\xi}{3\alpha_{e}}\pi(3+\pi^2)\right]^{3/2}$. Thus 
both the instabilities will have 
comparable growth rate when $\xi_{c}\sim 2^{2/3}\left[\frac{3\alpha_{e}}{2{\pi(3+\pi^2)}}\right]$. For 
$\xi<\xi_{c}$, chiral instability will dominate  else the Weibel 
instability will dominate.
In figure (1) we plot the dispersion relation given by Eq.(\ref{eq5}) as function
of $k_N$ for various values of $\xi$ and propagation angle $\theta_{n}$. $y$-axis
shows the $Re[\omega]$ and $Im[\omega]/\left(\frac{4\alpha_{e}^3\mu^3}{\pi^{4}m^2_{D}}\right)$ . Note 
that $Im[\omega]=\rho(k)$ and $Re[\omega] =0$. 
First note that when $\xi=0$  the unstable modes
could only be due to the chiral-imbalance.  The blue curves in fig.(1a,1b, 1e) depict this case. 
For the  sake of comparison
we have also plotted the pure Weibel modes by dropping the unity from Eq. (\ref{result:1}).
The green curves in fig(1a,1b,1e) represent this case. When $\xi\neq 0$ there is a contribution from both the 
instabilities and the condition for 
the instability can be written as
\begin{equation}
 \frac{\xi}{12}\left(1+3 cos2\theta_n\right)\left(\frac{\pi^2 m^2_D}{\mu^2\alpha_{e}^2}\right)
+k_N-k^2_N>0.
\label{condition1}
\end{equation}
Thus for a sufficiently large values of $k_N$ there is always a damping and this is consistent
with the findings of Weibel instability \cite{Weibel:1959, Fried:1959}. 
The above inequality can be solved rather easily and since $\frac{\pi^2m^2_D}{4\mu^2\alpha_{e}^2}\gg 1$
one can write the condition for the instability as
\begin{equation}
 0<k_N< 1+\frac{\xi}{3}\left(1+3 cos2\theta_n\right)\left(\frac{\pi^2m^2_D}{4\mu^2\alpha_{e}^2}\right).
\label{condition2}
\end{equation}

First one can notice from condition(\ref{condition2}) that
by increasing $\xi$ the contribution of Weibel instability
increases significantly when $\theta_n=0$. This is 
because $\frac{\pi^2m^2_D}{4\mu^2\alpha_{e}^2}\gg1$ and as 
$\xi$  becomes sufficiently large the Weibel instability terms start dominating over the terms 
due to the chiral-imbalance.
Further, we have already noted that the chiral instability occurs within the range $0<k_N<1$. From 
condition (\ref{condition2}) one can see that for small values of $\theta_n$, 
the range of the instability
can go beyond $k_N=1$. For example when $\theta_n=0$, 
the condition for the instability is $k_N<1+\frac{4}{3}\xi\left(\frac{\pi^2m^2_D}{4\mu^2\alpha_{e}^2}\right)$.
\begin{figure}[H]
\centering
\subfigure[]{\includegraphics[width= 8.0cm]{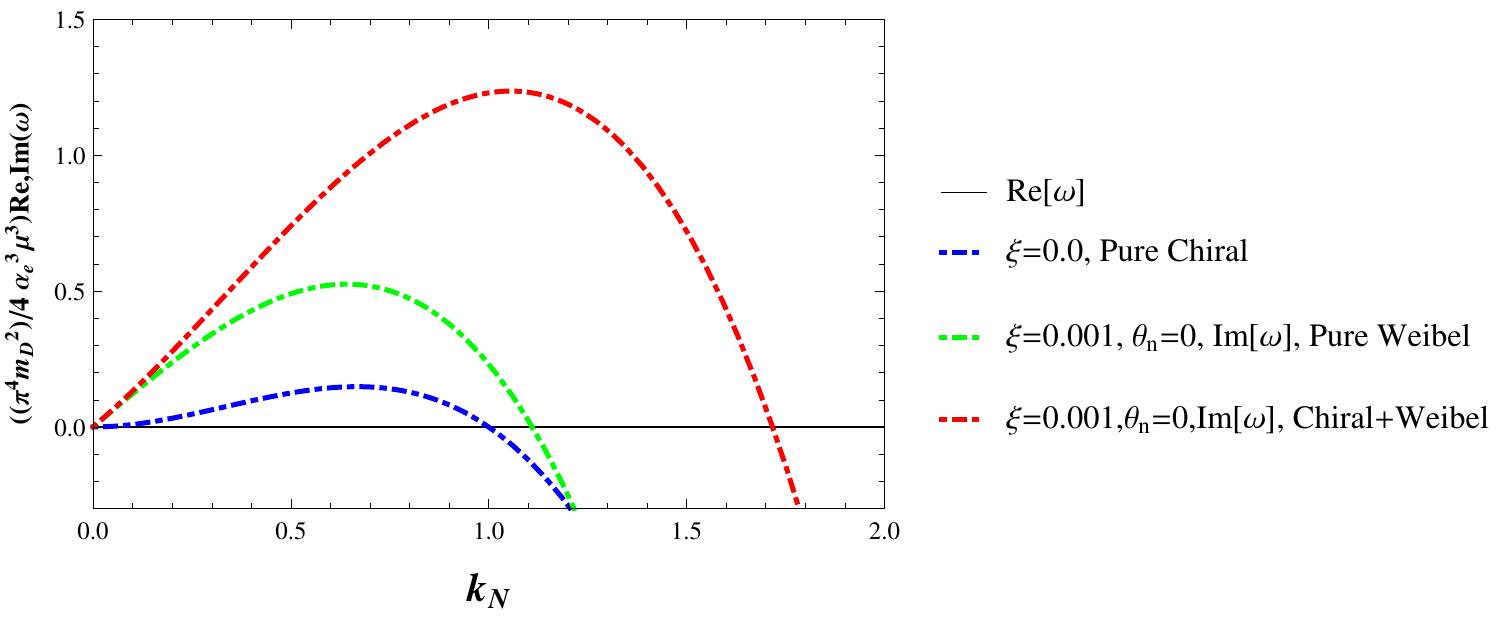}}\hspace{0.0cm}
\subfigure[]{\includegraphics[width= 8.0cm]{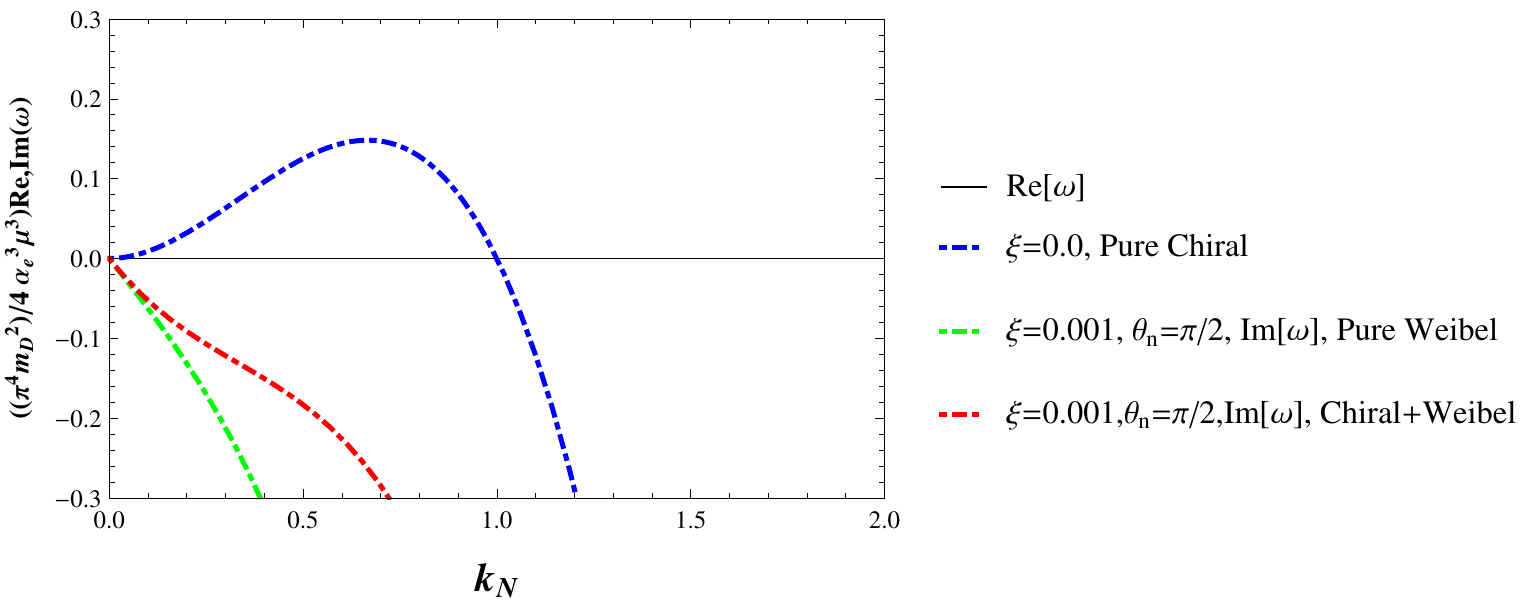}}\vspace{0.0cm}
\subfigure[]{\includegraphics[width= 8.0cm]{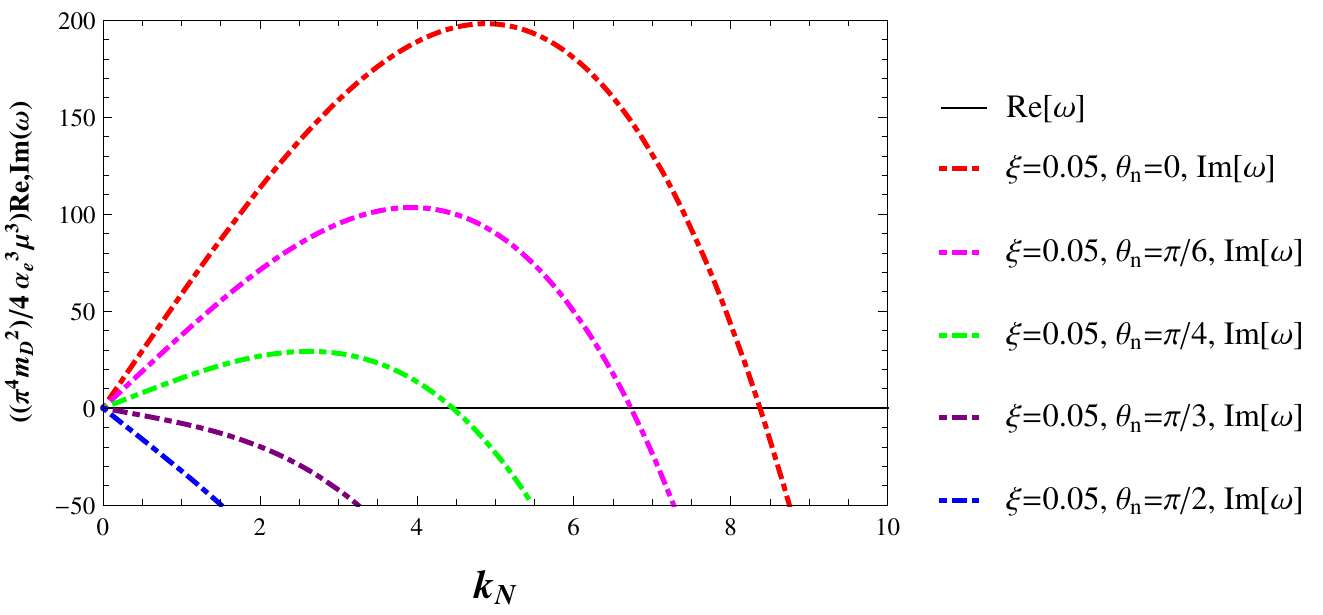}}\vspace{0.0cm}
\subfigure[]{\includegraphics[width= 8.0cm]{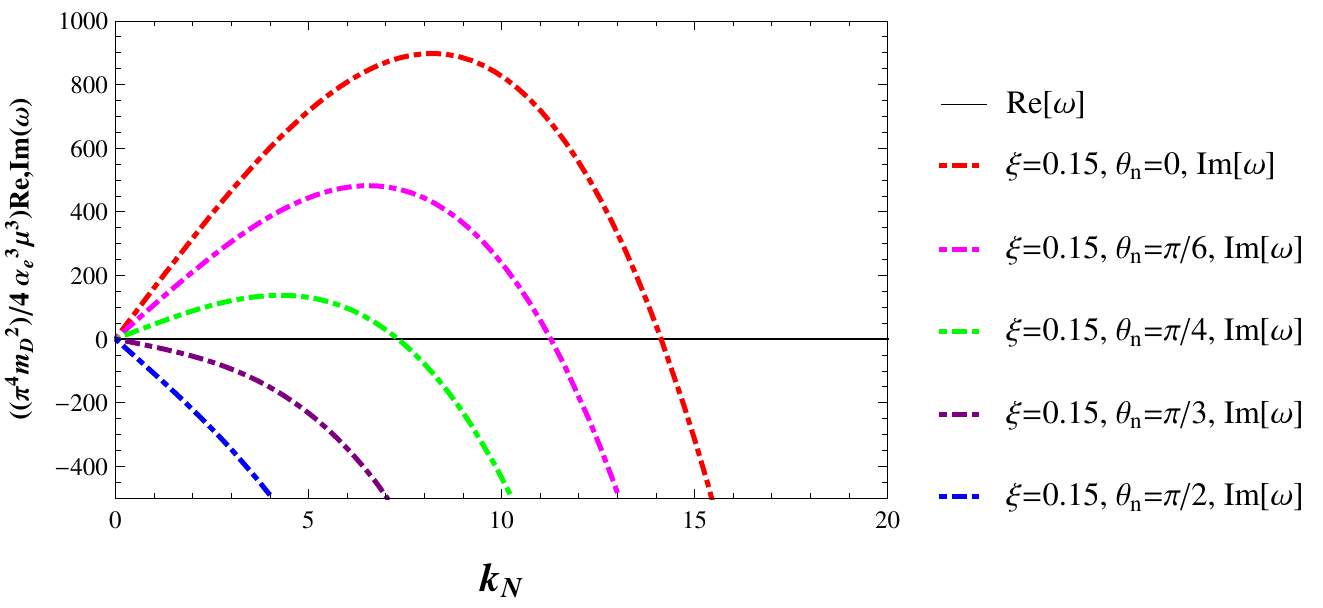}}\hspace{0.0cm}
\subfigure[]{\includegraphics[width= 8.0cm]{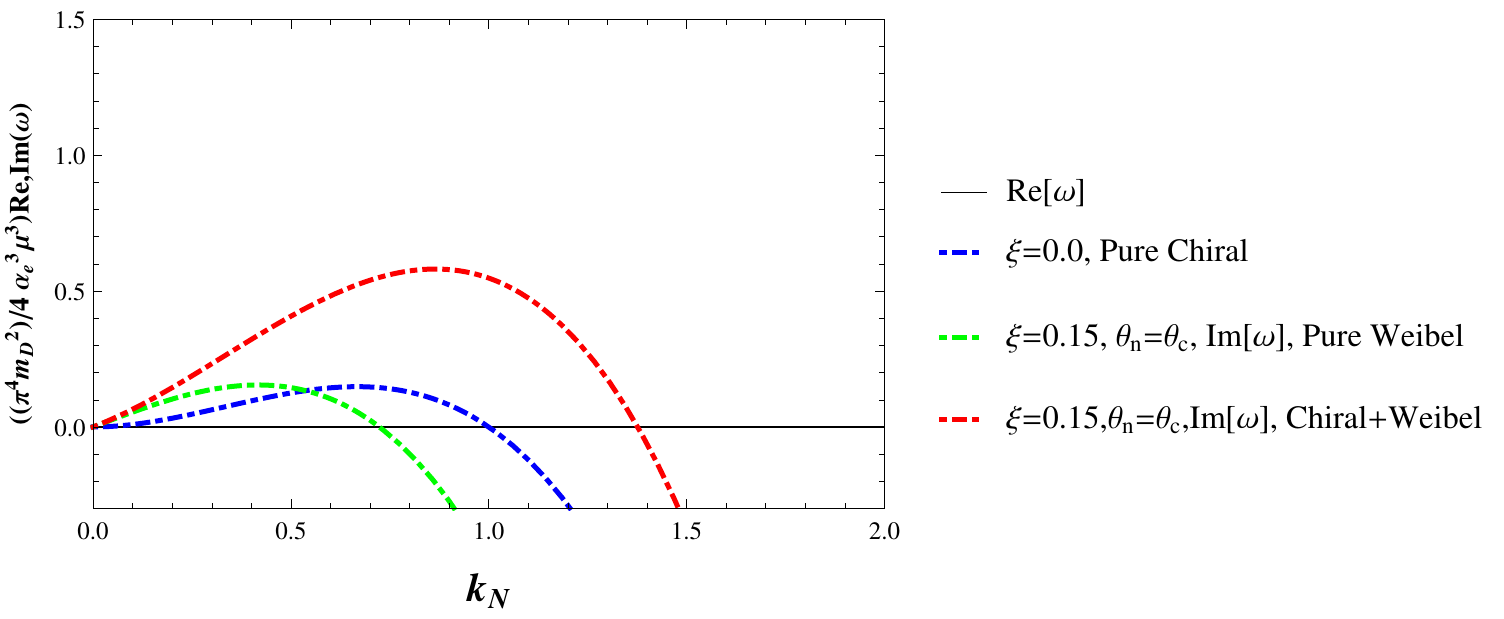}}\hspace{0.0cm}
  \caption
  { Shows plots of real and imaginary part of the dispersion relation. Here $\theta_{n}$ is the angle between 
  the wave vector $k$ and the anisotropy vector. Real part of dispersion relation is zero. Fig. (1a-1b) 
  show plots for three cases: (i) Pure chiral (no anisotropy), (ii) Pure Weibel (chiral chemical potential=0) and (iii) 
  When both chiral and Weibel instabilities are present. Fig. (1c-1d) represent the case when both the 
  instabilities are present but the anisotropy parameter varies at different values of $\theta_{n}$. Fig. (1e) represents the case when for a 
  particular value of $\theta_n\sim\theta_{c}$ both the instabilities have equal growth rates. Here frequency is normalized in unit
of $\omega/\left(\frac{4\alpha^3\mu^3}{\pi^{4}m^2_{D}}\right)$ and wave-number
$k$ by $k_{N}=\frac{\pi}{\mu\alpha}k$.} 
\label{fig[1]}
\end{figure}

For instance when $\theta_n=\pi/2$, range of the instability reduces from $k_N=1$ and it is
given  by $k_N<1-\frac{2}{3}\xi\left(\frac{\pi^2m^2_D}{4\mu^2\alpha_{e}^2}\right)$. 
This can be seen in the behavior of the plots of the dispersion relation shown in Figs.(1a-1b).
It is interesting to note that when the anisotropic parameter $\xi$, satisfies the 
condition $-1<\xi<0$, the conclusions about the range of the instabilities can
be altered and now the range of instability would reduce from $k_N=1$ for $\theta_n=0$ case
and it increases for $\theta_n=\pi/2$ case.
The negative value of $\xi$ signifies that the distribution function
in the momentum space is stretched in the direction of the anisotropy vector ${\bf n}$.
In fig.(1a) and fig (1b) the red curve respectively show the instance when both the instability occurs simultaneously for 
$\theta_n=0$ and $\theta_n=\pi/2$ cases for $\xi=0$. Figure(1a) shows that for $\theta_n=0$ the combined growth rate of the instability is
significantly higher than the pure chiral and the pure Weibel cases. In this case the range of the instability also increases
in comparison with the pure cases. However for $\theta_n=\pi/2$, the Weibel instability is absent and the combined mode
also is damped as can be seen from fig.(1b). Figures (1c and 1d) show that range and the growth rate of the combined 
instability sensitively depends on $\xi$.  As $\xi$ value increases the Weibel instability dominates over the chiral 
instability.  However there exists a some critical value for $\theta_{c}
=\frac{1}{2}\cos^{-1}\left[\left(\frac{2}{27}\right)^{2/3}\frac{12\mu^2\alpha^2}{\xi\pi^2m_{D}^2}-\frac{1}{3}\right]$ for 
a given $\xi$, where both the instabilities
contribute equally.  This case is depicted in fig.(1e).

\section{Stream passing through chiral plasma}

Another important class of the problem that deals with anisotropic situation is
the case when a stream of particles moving in a thermalized back-ground plasma \cite{Krall}. 
The stream can loose its energy and momentum by interacting with the plasma. This kind of problem have
applications in variety of fields including quark-gluon plasma 
\cite{Nieves:2000, Ng:2006, Manuel:2008, Manuel:2006, Pisarski:97}.
In the present work for background plasma we consider the background distribution function 
in momentum space $n^{0}_{\mathbf{p}}$ to be isotropic. We write 
$n^{0}_{\mathbf{p}}=n^{0(0)}_{\mathbf{p}}+e n^{0(\epsilon\delta)}_{\mathbf{p}}$ where,   
$n^{0(0)}_{\mathbf{p}}=\frac{1}{[e^{(p-\mu)/T}+1]}$ and 
$n^{0(\epsilon\delta)}_{\mathbf{p}}=\left(\frac{{\mathbf{B}\cdot\mathbf{v}}}{2 p T}\right)
\frac{e^{p-\mu)/T}}{[e^{(p-\mu)/T}+1]^2}$ which is same as considered in Ref. \cite{Akamatsu:2013}.
Note that we can obtain this form of the distribution function from the equilibrium distribution function
considered in the previous section by setting the anisotropy parameter $\xi=0$.
Next consider a jet or stream of chiral fermions traveling in the background plasma with the spatial
component of the four-velocity ${\mathbf u}=u^0{\mathbf v_{st}}$, where $u^0$ is temporal component of
the four velocity of the stream and ${\bf v_{st}}$ is the three vector of the stream velocity. 
${\bf v_{st}}$ is same and constant for all the particles in the jet. In this
case we consider the following equilibrium distribution function for the jet:   
\begin{equation}
n^{0}_{\mathbf{p}}=(2\pi)^3\bar{n}{u^{0}}\delta^{3}(\mathbf{p}-\Lambda\mathbf{u})(1+e\mathbf{B.\Omega_p})
\label{jetdistro}
\end{equation}
\noindent where, $\bar{n}$ describes the density of the jet particles and it is considered to be constant.
Here $\Lambda$ is the scale of energy of the jet. The term with ${e\bf B\cdot \Omega_p}$
 is Berry curvature correction to the distribution function of the stream and it is $O(\epsilon\delta)$.  
 
 Now one can consider the perturbations in the distribution function of
the background plasma and the jet. The self-energy expression for the background plasma
can be simply obtained by taking $\xi\rightarrow 0$ limit from 
 Eqs.(\ref{selfenergyoepsilon},\ref{selfenergyoepsilondelta}) as, 
\begin{eqnarray}
 {\Pi^{ij}_{+}(K)}&=&
\alpha P^{ij}_L+\beta  P^{ij}_T\\
 {\Pi^{ij}_{-}(K)}&=&
\lambda P^{ij}_A
\end{eqnarray}
where $\alpha=\Pi_{T}$, 
$\beta=\frac{\omega^2}{k^2}\Pi_{L}$, $\lambda=-\frac{\Pi_{A}}{2}$.  $\Pi_{T}$, $\Pi_{L}$ and $\Pi_{A}$ are given 
by Eq.(\ref{isotropic:eq}).

For the case of the jet anomalous Hall-current term in general can be non-zero.
Due to the presence the delta function in the distribution function in Eq. (\ref{jetdistro})
it is rather easy to calculate the expressions for parity-even $\Pi^{ij}_{+st} $ and parity-odd 
$\Pi^{ij}_{-st}$ parts of the self-energy tensor associated with the jet and they are given
below:
\begin{equation}
 {\Pi^{ij}_{+st}(K)}=\omega^2_{st}\left[\delta^{ij}+
 \frac{k^{i}{v^j_{st}}+k^{j}{v^i_{st}}}{\omega-\mathbf{k.v_{st}}}-
 \frac{(\omega^2-\mathbf{k^2})v^i_{st}v^j_{st}}{(\omega-\mathbf{k.v_{st}})^2}\right],
\label{jetselfenergyoepsilon}
\end{equation}
\begin{eqnarray}
{\Pi^{im}_{-st}(K)}&=&\frac{i\omega{\epsilon}^{jlm}k^{l}v^{j}_{st}v^{i}_{st}\omega^2_{st}}{2\Lambda{u^0}(\omega-
 \mathbf{k\cdot{v_{st}}})}-
 \frac{i{\epsilon}^{ilm}k^{l}(1-2 v^2_{st})\omega^2_{st}}{2\Lambda{u^0}}-
 \frac{i\omega{\epsilon}^{iml}v^{l}_{st}\omega^2_{st}}{2\Lambda{u^0}}\nonumber\\
  &+&\frac{i{\epsilon}^{ijl}k^{l}\omega^2_{st}}{2\Lambda{u^0}}\left[\delta^{jm}+
 \frac{k^{j}{v^m_{st}}+k^{m}{v^j_{st}}}{\omega-\mathbf{k.v_{st}}}-
 \frac{(\omega^2-\mathbf{k^2})v^j_{st}v^m_{st}}{(\omega-\mathbf{k.v_{st}})^2}-
 \frac{\omega{v^j_{st}v^m_{st}}}{(\omega-\mathbf{k\cdot v_{st}})}\right],
\label{jetselfenergyoepsilondelta}
\end{eqnarray}
where $\omega^2_{st}=\frac{\bar{n} e^2}{2 \Lambda}$. Note that third term on the right 
hand side of equation (\ref{jetselfenergyoepsilondelta})
is due to the anomalous Hall-current. The total self-energy of the system can be obtained by
adding the contributions from the background plasma and the jet:
\begin{equation}
 {\Pi^{ij}(K)}= {\Pi^{ij}_{+}(K)}+ {\Pi^{ij}_{-}(K)}+{\Pi^{ij}_{+st}(K)}+{\Pi^{ij}_{-st}(K)}
\label{pitotal:1}
\end{equation}
Next, we use Eqs.(\ref{invprop},\ref{pitotal:1}) to analyze the modes in a jet-plasma system.

\subsection{Study of collective modes of the system of chiral plasma with a stream}

In order to analyze the collective mode one can evaluate determinant of $[\Delta^{-1}(K)]^{ij}$ :
\begin{equation}
det[[\Delta^{-1}(K)]^{ij}]=det[({k^2}-{\omega}^{2})\delta^{ij}-k^{i}k^{j}+\Pi^{ij}(K)]=0. \label{findingpoles}
\end{equation}
 In what follows we choose the streaming velocity ${\bf v_{st}}$ in $z-$direction only and the wave propagation vector
${\bf k}$ has a component 
in a direction
parallel to ${\bf v_{st}}$ i.e. $k_z$.

\subsubsection{When $\bf{k}$ parallel to $\bf{v_{st}}$}

 In this case ${\bf k}\cdot{\bf v_{st}}=kv_{3st}$, solution of equation (\ref{findingpoles}) gives the following
dispersion relation: 
\begin{equation}
\left(4\bar{\Lambda}^2(k^2-\omega^2+\alpha+\omega^2_{st})-(2\bar{\Lambda}\lambda+(3 k-v_{3st}(\omega+2kv_{3st}))\omega^2_{st})^2\right)
\left(\beta+\omega^2\left(-1-\frac{(-1+v^2_{3st})\omega^2_{st}}{(\omega-kv_{3st})^2}\right)\right)=0,
\end{equation}
where $\bar{\Lambda}=\Lambda/(1-v^2_{3st})^{1/2}$. Thus there exists two separate branches for the mode of propagation, 
\begin{eqnarray}
\left(\beta+\omega^2\left(-1-\frac{(-1+v^2_{3st})\omega^2_{st}}{(\omega-kv_{3st})^2}\right)\right)=0,\label{cristinaseqn}\\
\left(4\bar{\Lambda}^2(k^2-\omega^2+\alpha+\omega^2_{st})^2-(2\bar{\Lambda}\lambda+(3 k-v_{3st}
(\omega+2kv_{3st}))\omega^2_{st})^2\right)=0.\label{new}
\end{eqnarray}
Eq.(\ref{cristinaseqn}) is exactly same as discussed in Ref.\cite{Manuel:2008}
and it solutions will not be discussed here. However, interestingly 
this branch does not get any correction due to parity-odd effect considered in this work.
 Eq.(\ref{new}) is a new branch of the dispersion relation arising entirely 
due to the parity odd effect. Next, we analyze this new branch in the quasi 
static limit $|{\omega}|<<k$, one can write 
\begin{eqnarray}
\alpha_{|{\omega}|<<k}\approx-i\frac{\pi}{4}\frac{\omega}{k}m^{2}_{D},\nonumber\\
\beta_{|{\omega}|<<k}\approx-m^{2}_{D}\frac{\omega^2}{k^2},\nonumber\\
\lambda_{|{\omega}|<<k}\approx\frac{\mu k e^2}{4\pi^2}.\label{quasistationarylimit}
\end{eqnarray}
From Eq.(\ref{new},\ref{quasistationarylimit}) and using $\omega=A+iB$ where $A$ and $B$ are real and imaginary 
part of $\omega$ one can obtain:
\begin{equation}
B(k)=+\frac{\frac{4 \alpha \mu k^2}{\pi^2 m^2_D}\left[1-\frac{\pi k}{\alpha \mu}-\frac{\pi \omega^2_{st}}{\alpha \mu k}
\left(1-\frac{3 (1-v^2_{3st})^{1/2}}{2}\frac{k}{\Lambda}+v^2_{3st}(1-v^2_{3st})^{1/2}\frac{k}{\Lambda}\right)\right]}
{\left(1+\left(\frac{2 k v_{3st} (1-v^2_{3st})^{1/2}
\omega^2_{st}}{\pi m^2_{D} \Lambda}\right)^2\right)}. \label{imaginarypartofomega}
\end{equation}
In the above equation first term inside the square bracket is 
arising due to the chiral-imbalance in the plasma.
The third term $-\frac{\pi \omega^2_{st}}{\alpha \mu k}
\left(1-\frac{3 (1-v^2_{3st})^{1/2}}{2}\frac{k}{\Lambda}+v^2_{3st}(1-v^2_{3st})^{1/2}\frac{k}{\Lambda}\right)$ 
is the effect of streaming. Note 
that the terms with $\frac{k}{\Lambda}$ are the parity violation or chiral imbalance contribution to the 
stream. For the case when $\omega_{st}=0$, one recovers the chiral-imbalance instability
discussed in Ref.(\cite{Akamatsu:2013}). For
$\omega_{st}\neq0$ one can see three terms in small bracket of numerator in Eq.(\ref{imaginarypartofomega}) 
competing with each other may give overall positive or negative contribution to
the instability depending on the values $k$, $v_{3st}$ and $\Lambda$.
It is convenient to define the total plasma frequency $\omega_t$ using $\omega^2_t=\omega^2_p+\omega^2_{st}$,
where $\omega_p$ is the plasma frequency and ($\omega^2_{p}=\frac{m^2_D}{3}$). 
Using this we introduce normalize frequency $\omega_1=\omega/\omega_{t}$
and wave-number $k_1=k/\omega_{t}$. Further we have the following parameters:
 $b=\omega^2_{st}/\omega^2_{t}$, 
$\mu_1=\mu/\omega_{t}$, $\Lambda_1=\Lambda/\omega_{t}$. 
It should be noted that parameters $\mu_1$  arises due to the parity-odd effect and
it was not there in Ref.\cite{Manuel:2008}. 
For a finite temperature plasma when $\mu\sim T$,
 one can have $\mu_1=3\pi\left(\frac{2(1-b)}{3+\pi^2}\right)^{1/2}$. For Heavy ion collisions typical value
of $\Lambda$ can be taken $4-100$ GeV/c\cite{Aad:2011,Abelev:2006}  
and $\Lambda_{1}\approx30$ or greater depending upon $\Lambda$.  However for Weyl metals values of $\Lambda$
can be much lower and it may have different values.  
One can now rewrite Eq.(\ref{imaginarypartofomega}) as follows:

\begin{equation}
B_1(k)=+\frac{\frac{4 \alpha \mu_1 k^2_1}{3 \pi^2 (1-b)}\left[1-\frac{\pi k_1}{\mu_1 \alpha}-\frac{\pi b}{\alpha \mu_1 k_1}
\left(1-\frac{3 (1-v^2_{3st})^{1/2}}{2}\frac{k_1}{\Lambda_1}+v^2_{3st}(1-v^2_{3st})^{1/2}\frac{k_1}{\Lambda_1}\right)\right]}
{\left(1+\left(\frac{2 k_1 
v_{3st} (1-v^2_{3st})^{1/2}b}{3 \pi (1-b) \Lambda_1}\right)^2\right)}.\label{imaginarypartofomega1}
\end{equation}
In Fig.(2) we have shown the plots of $Im[\omega_1]$ i.e. $B_1(k)=B(k)/\omega_t$  versus $k_1$ for various
values of parameters $b$, stream-velocity $v_{3st}$ and $\Lambda_1$.
Fig.2(a) shows the plots of $B_1(k)$ as a function of $k_1$ for different values of $b$
while $v_{3st}$ and $\Lambda_1$ are kept fixed at $v_{3st}=0.9$ and $\Lambda_1=30$.
The 
case with $b=0$ corresponds to the case when there is no stream and the dispersion 
relation gives the same instability for the plasma background considered in Ref.(\cite{Akamatsu:2013}). 
But 
by increasing  $b$ the background plasma instability
is reduced because the
term with factor $-\frac{\pi b}{\alpha \mu_1 k_1}$ in Eq.(\ref{imaginarypartofomega1})
gives a strong negative contribution to the instability. Keeping $b=0.01$ and $\Lambda_1$
in the similar ballpark as in Ref.\cite{Manuel:2008} (relevant for a QGP) and $v_{3st}=0.9$ can strongly
suppress the back-ground instability. 
Fig.(2b) shows how the plots varies with different
values of $\Lambda_1$ while we have kept parameters $b$ and $v_{3st}$ fixed at $b=0.01$ and $v_{3st}=0.9$. 
Note that case with $b=0$ is shown for just making a comparison with the background plasma
instability.
In this case one can see that when $\Lambda_1\ll 1$ is the instability
is enhanced compared to the background plasma case with $b=0$. This is arising
because of the parity-odd contribution to the self-energy coming from the jet.
The reason for this is, the term  with coefficient $1/\Lambda_1$ dominates in Eq.(\ref{imaginarypartofomega1})
and make a strong positive contribution to the instability.
As we increase the value of $\Lambda_1$ the instability is strongly suppressed.
Fig.2(c) shows the case when $b$ and $\Lambda_1$ are kept fixed at $b=0.03$ and $\Lambda_1=0.1$
while parameter $v_{3st}$  varies.
One can see here that parity-odd terms in jet can enhance the instability around $v_{3st}=0.3$.
But the contribution from the parity-odd terms in jet reduces significantly as
$v_{3st}\rightarrow 1$.
\begin{figure}[H]
 \centering
\subfigure[]{\includegraphics[width= 10.0cm]{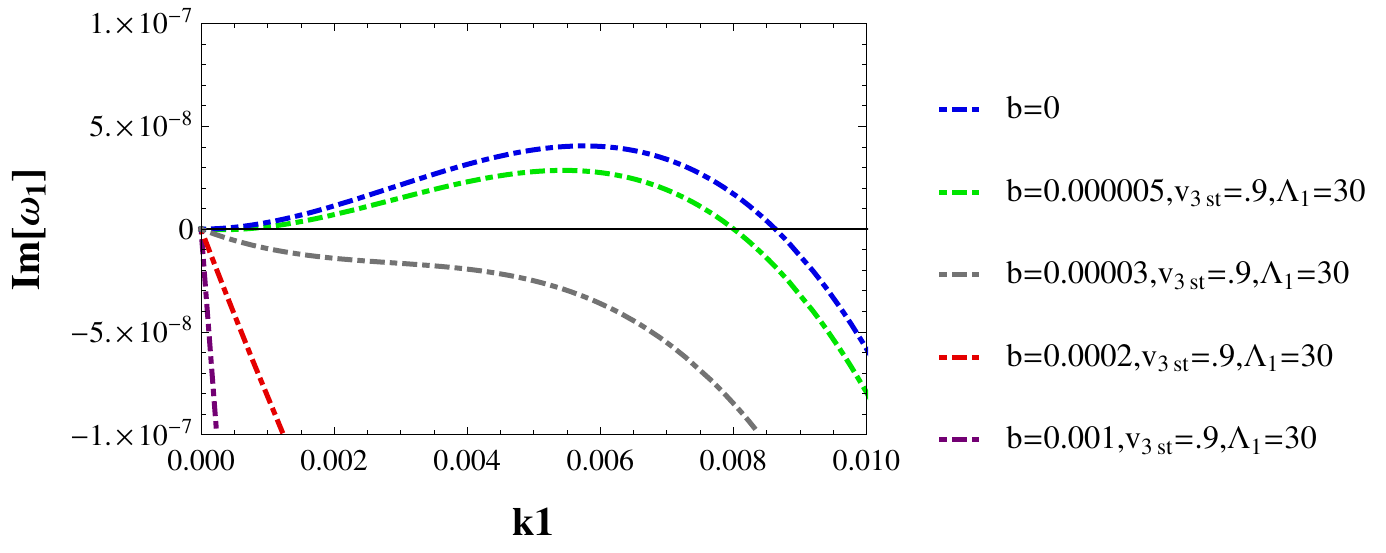}}\hspace{0.0cm}
\subfigure[]{\includegraphics[width= 9.0cm]{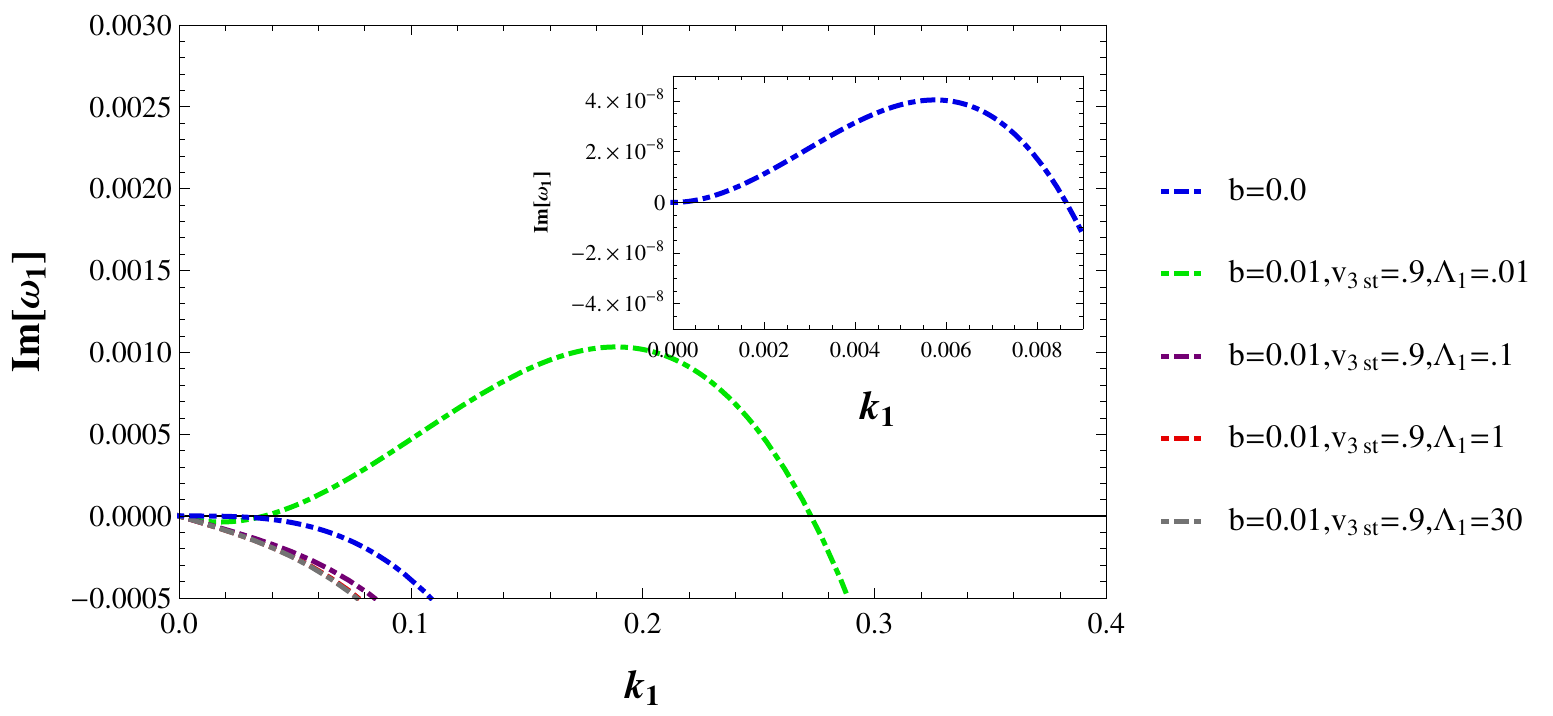}}\vspace{0.0cm}
\subfigure[]{\includegraphics[width= 9.0cm]{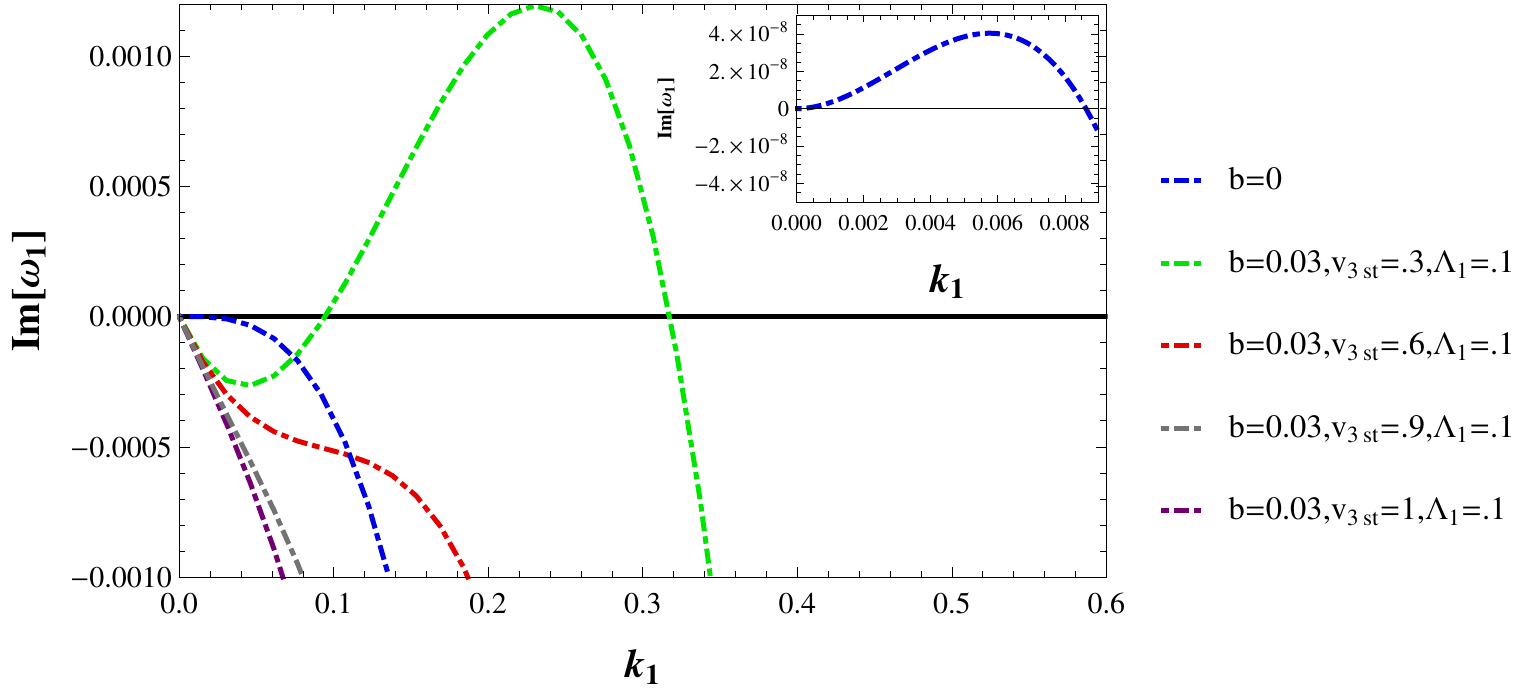}}\vspace{0.0cm}
\caption
   { show plots of dispersion relation of the instability in a chiral-plasma
background with a stream for the situation when the wave-vector $k_1$ propagating in
the direction parallel to the stream velocity $v_{st}$. The $b=0$ corresponds to the situation 
when there is no stream in the background  plasma. 
Fig.(2a) shows how the instability varies for different values of $b$ while
the stream velocity $v_{3st}=0.9$ and $\Lambda_1=30$. Fig.(2b) shows that for a given
values of $b$ and $v_{3st}$, the parity-odd terms in the jet self-energy can enhance the instability.
Fig.(2c) shows the dependence of the instability on the stream velocity for given
values of $b$ and $\Lambda_1$. Inset Figures in Fig. (2b,2c) shows the instability for b=0 case with better 
resolution.}
\label{fig[2]}
\end{figure}

\subsubsection{When $\bf{k}$ perpendicular to $\bf{v_{st}}$}

In this case $\bf{k}\cdot\bf{v_{st}}=0$. By choosing $\mathbf{k}$ to be in x-direction and $\mathbf{v}$ in 
z-direction, Eq.({\ref{findingpoles}}) gives following dispersion relation 
in limit $|\omega|<<k$ :
\begin{align}
\left(\lambda^2_1-(k^2_1+\alpha_1+b)^2\right)(-\omega^2_1+\beta_1+b)-(k^2_1+\alpha_1+b)\left(\left(-1+
\frac{\beta_1}{\omega^2_1}\right)b k^2_1 v^2_3\right)\nonumber\\
+6\left(\frac{k_1(1-v^2_3)^{1/2}}{2\Lambda_1}\right)\lambda_1 b \Bigg((\omega^2_1-\beta_1)\left(1-\frac{k^2_1v^2_3}{6 \omega^2_1}\right)+
\left(1-\frac{v^2_3}{2}\right)b\Bigg)\nonumber\\+ 
\Big(\frac{k_1(1-v^2_3)^{1/2}}{2\Lambda_1}\Big)^2b^2\Bigg(9\left(\omega^2_1\Big(1-\frac{v^2_3}{2}\Big)-\beta_1\left(1-
\frac{k^2_1v^2_3}{3 \omega^2_1}\Big(1-\frac{v^2_3}{3}\Big)\right)+\frac{k^2_1v^4_3}{9}-b\right)\nonumber\\+v^2_3(-2 k^2_1+\alpha_1+7 b)\Bigg)
=0\label{finaldispersion}
\end{align}
where, $\lambda_1=\frac{\mu_1 k_1\alpha}{\pi}\left(1 + i\frac{\pi}{2}\frac{\omega_1}{k_1}\right)$, 
$\alpha_1=-i\frac{3 \pi (1-b)}{4}\frac{\omega_1}{k_1}$, 
$\beta_1=-3(1-b)\frac{\omega^2_1}{k^2_1}$. 
Note that here we have introduced same normalized variable and the parameters defined for 
$\mathbf{k}\parallel \mathbf{v_{st}}$ case. It is very clear from the above equation that for $b=0$
 one gets the dispersion relation for the background-plasma (\cite{Akamatsu:2013}). 
If the terms with parameter $b$ are kept and the terms with $\mu_1$ and $\frac{1}{\Lambda_1}$ are dropped,
one can obtain the standard dispersion relation obtained in [e.g. see Ref.\cite{Manuel:2008}] for
jet-plasma system for a parity-even case.

In Fig.(3) we plot the $\omega_1$ calculated using Eq.(\ref{finaldispersion}) as function of
$k_1$ for different values of parameters $b$, $\Lambda_1$ and $v_{3st}$. 
In Fig. 3(a) we have plotted the positive roots of imaginary part 
of $\omega_1$ with respect to $k_1$. The comparison of the root with $b=0.01$, $\Lambda_1=30$ and
$v_{3st}=0.06$ is made with the parity-even plasma\cite{Manuel:2008} 
and the no-jet case $b=0$ with \cite{Akamatsu:2013}. 
Thus from the Fig. 3(a) it is clear that the presence of parity-violation effect in the stream enhances the 
instability. Next, Fig 3(b) we have increased the stream velocity to $v_{3st}=0.065$ from its value
$v_{3st}=0.06$ and all other condition remains same between the two figures. In this
case also the
instability is enhanced due to increased velocity of jet and the parity-odd effect. Note that in comparison
with the instability in no-jet case the finite jet has a much stronger instability.
In Fig. 3(c) we have shown how the streaming instability will change by changing
parameter $\Lambda_1$. For $\Lambda_1 <1$, the parity-odd terms can enhance the 
instability provided stream-velocity remains sufficiently small.
Next, in Fig. 3(d) we have shown the 
variation of the instability by changing
the parameter $b$ while parameters $\Lambda_1$ and $v_{3st}$ are kept fixed.
The instability decreases as we decreases the value of $b$ 
but the three different curve covers the different $k_1$ values. One can see that 
when $b=0.001$ the system can be unstable for larger $k_1$ value as compared to cases when $b=0.02$ and $0.1$.
Further, we would like to note that when the stream velocity increases  and approach unity,
the parity-odd contribution from the jet becomes negligible and the contribution
to the instability from the parity-odd background remains much weaker. In this limit
the parity-even contribution in Ref.\cite{Manuel:2008} can remain unaltered.
\begin{figure}[H]
 \centering
\subfigure[]{\includegraphics[width= 10.0cm]{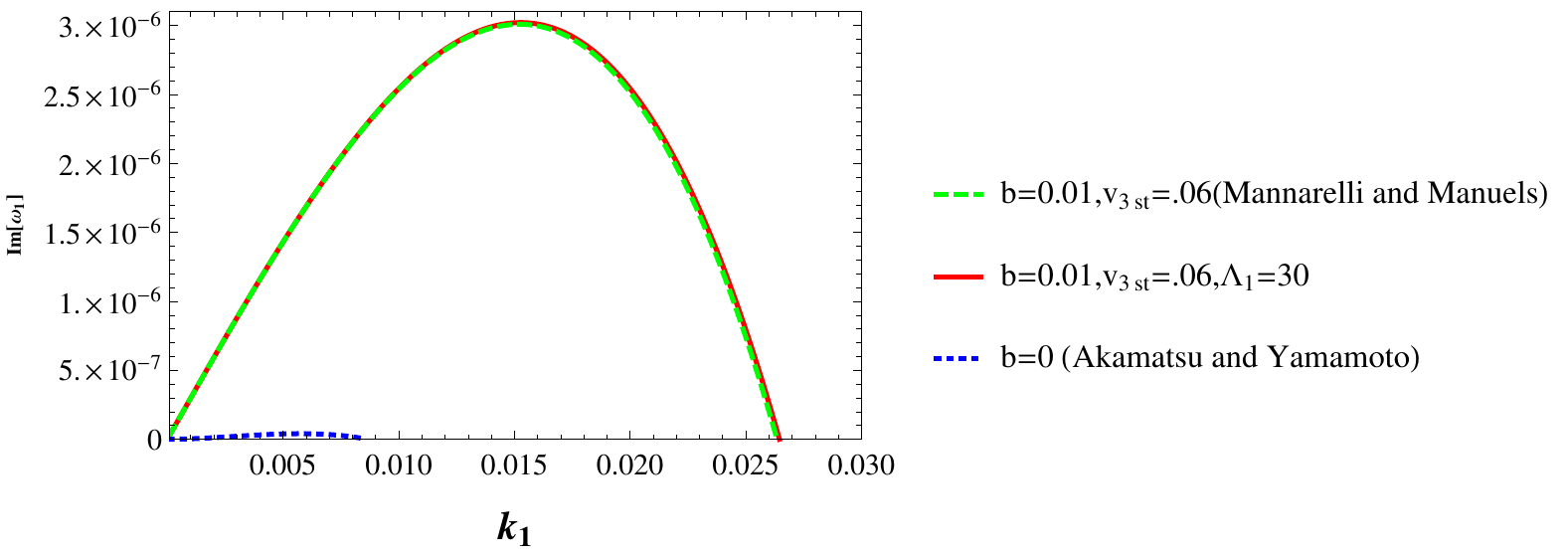}}\hspace{0.0cm}
\subfigure[]{\includegraphics[width= 10.0cm]{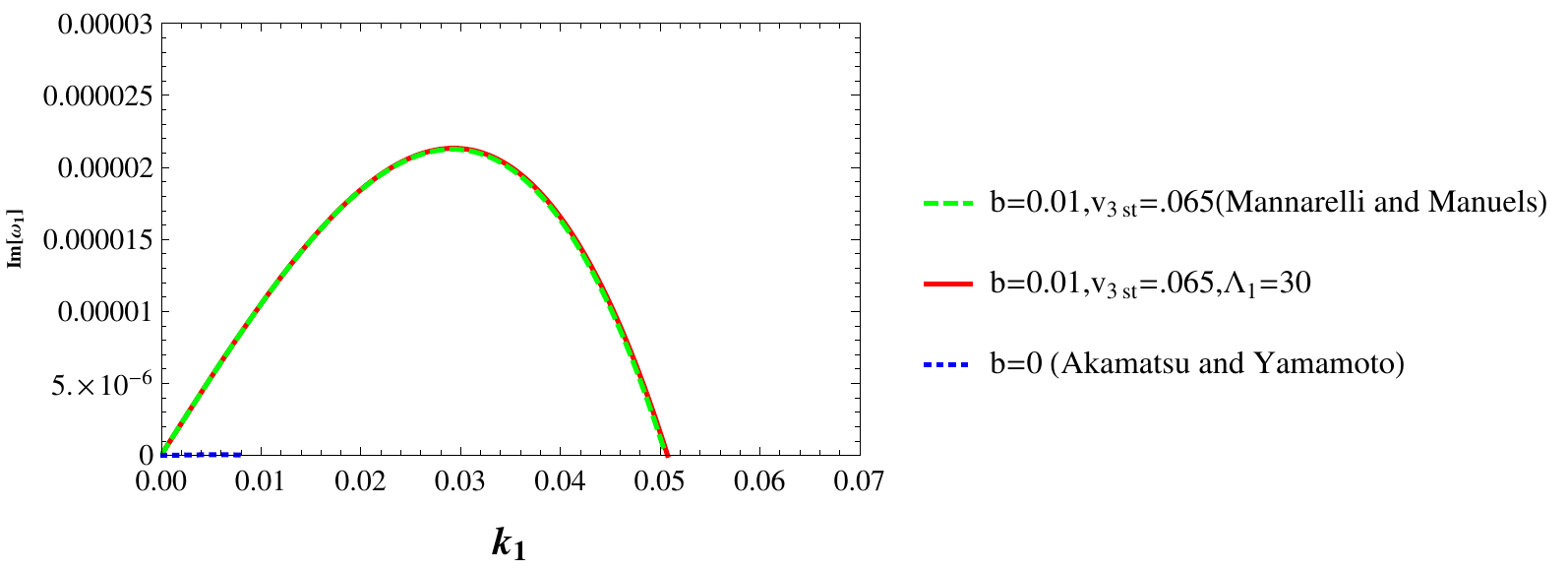}}\vspace{0.0cm}
\subfigure[]{\includegraphics[width= 9.0cm]{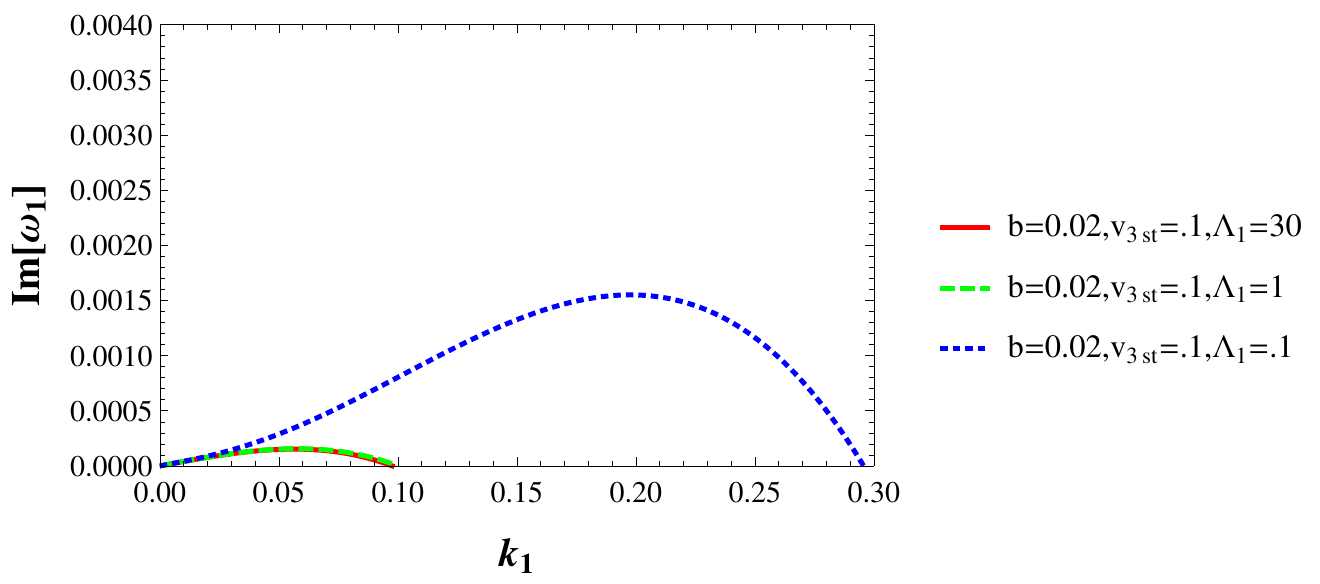}}\hspace{0.0cm}
\subfigure[]{\includegraphics[width= 9.0cm]{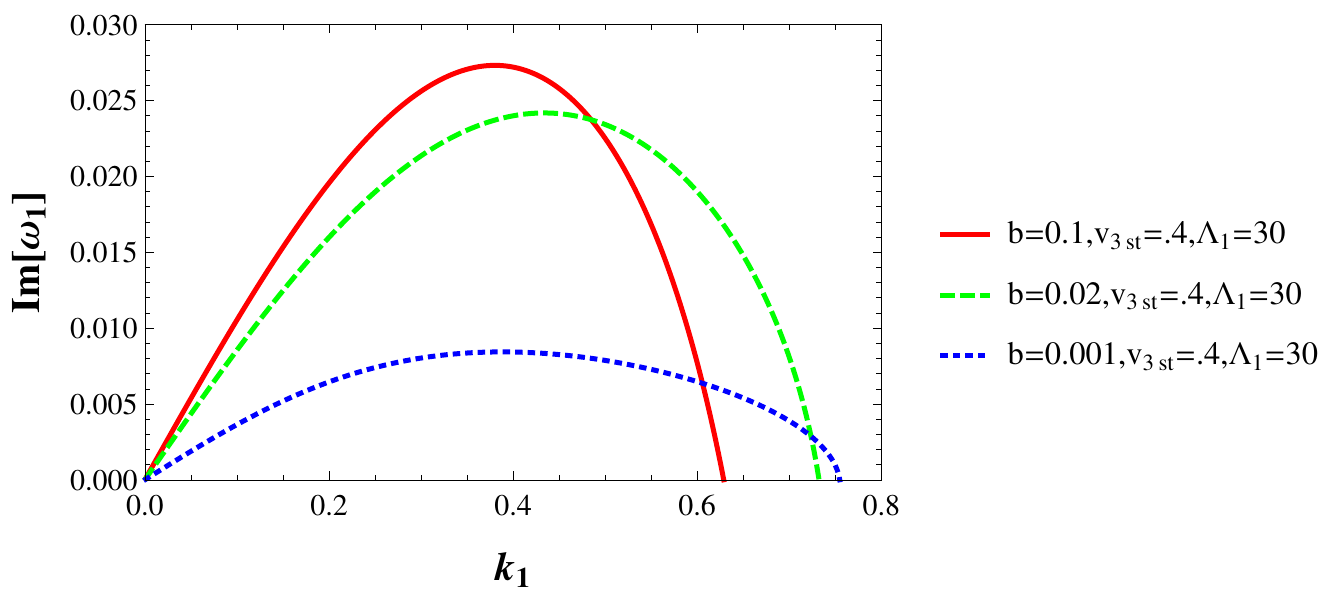}}\vspace{0.0cm}
\caption{show plots of dispersion relation of instabilities in a chiral plasma with a stream passing through 
   it when $\mathbf{k_1}$ perpendicular to $\mathbf{v_{st}}$. Fig. 3(a) shows a comparison in the instabilities when a 
stream with parameters $b=0.01$ and $v_{3st}=0.06$, $\Lambda_1=30$ passing through chiral plasma [red (solid) curve] 
to the cases, when there is no streaming i.e. $b=0$[blue (dotted) curve] and when there is stream 
with $b=0.01$, $v_{3st}=0.06$ passing through parity even plasma [green (dashed) curve]. 
Fig. 3(b) shows the same comparison at higher 
stream velocity keeping other parameter same as in case of Fig. 3(a). Fig. 3(c) shows the effect on instability by changing the 
parameter $\Lambda_1$ keeping parameters $b=0.02$ and $v_{3st}=0.1$ fixed. Fig. 3(d) shows the effect on instability by changing 
parameter $b$ keeping parameters $v_{3st}=0.4$ and $\Lambda_1=30$ fixed.}
\label{fig[3]}
\end{figure}  

\section{Summary and Conclusions}
 
 We have studied collective modes in anisotropic chiral plasmas. 
In particular we have considered two cases of the instabilities
in anisotropic plasma namely Weibel instability and jet-plasma interaction.
We have shown that even for small values of the anisotropy parameter $\xi\ll 1$,
the range and the magnitude of the chiral-imbalance instability is
strongly modified. For $\xi>0$, the growth rate and the range
increases significantly when the wave-vector $k$ is in the direction parallel to the anisotropy 
vector $\mathbf{n}$. The instability can become weaker when $k$ is in the direction
perpendicular to $\mathbf{n}$. In this case modes are strongly damped
when one increases value of $\xi$. 

 We have also studied the dispersion relation of a jet-plasma system with parity-odd
effect. We have shown that for the case when the wave-vector is in direction parallel 
to the stream velocity of the jet there can be two branches of the dispersion relation.
The standard branch that could arise in a parity-even plasma and the new branch that
is arising solely due to the parity odd effects of the chiral plasma. The standard branch
does not have any correction due to the Berry-curvature terms. For the new branch we have
shown that the chiral-imbalance instability is suppressed when the stream frequency
$\omega_{st}$ or parameter $b$ increases. Further, if the jet energy scale $\Lambda$
is much lower than that found in the heavy ion-collision, the parity-odd effect in
the stream can enhance the chiral-imbalance instability. Such lower values of
$\Lambda$ may be relevant for Weyl metals. However for the parameters of the jets in
the ballpark of relativistic heavy-ion collisions the chiral-imbalance instability
is strongly suppressed. For the case when the wave-vector is in direction  perpendicular
to the stream-velocity we have shown that the parity-odd effect can strongly enhance 
the streaming-instability when the stream velocity is small. However, when the 
the stream velocity become large the enhancement to the instability due to the parity-odd
effect become very small. We hope that the results presented here can be applicable
to the relativistic heavy-ion collisions and Weyl metals.  
 
{\bf{Acknowledgement:}} Our sincere thank to the referee whose comments has 
helped us in improving our presentations significantly.


\begin{thebibliography}{99}


\bibitem{Landau_kinetics}
 L.~D.~Landau and E.~M.~Lifshitz, {\it Physical Kinetics}
(Pergamon, New York, 1981).
%
\bibitem{Son:2009tf}
  D.~T.~Son and P.~Surowka,
  Phys.\ Rev.\ Lett.\  {\bf 103}, 191601 (2009)  [arXiv:0906.5044 [hep-th]].  

\bibitem{Banerjee:2012iz} 
  N.~Banerjee, J.~Bhattacharya, S.~Bhattacharyya, S.~Jain, S.~Minwalla, and T.~Sharma,
  JHEP {\bf 1209}, 046 (2012)
  [arXiv:1203.3544 [hep-th]].
  
\bibitem{Jensen:2012jy} 
  K.~Jensen,
  Phys.\ Rev.\ D {\bf 85}, 125017 (2012)
  [arXiv:1203.3599 [hep-th]].

\bibitem{Kharzeev:2010gr} 
  D.~E.~Kharzeev and D.~T.~Son,
  Phys.\ Rev.\ Lett.\  {\bf 106}, 062301 (2011).  [arXiv:1010.0038 [hep-ph]].  
\bibitem{Son:2012wh} 
  D.~T.~Son and N.~Yamamoto,
  Phys.\ Rev.\ Lett.\  {\bf 109}, 181602 (2012).
  [arXiv:1203.2697 [cond-mat.mes-hall]].
%
\bibitem{Zahed:2012}
I. Zahed, Phys. Rev. Lett. {\bf 109}, 091603 (2012).
%
\bibitem{Stephanov:2012}
M. A. Stephanov and Y. yin, Phys. Rev. Lett. {\bf 109}, 162001 (2012).\\
arXiv:1207.0747[hep-th]
%
\bibitem{Chen:2013}
Jiunn-Wei Chen, Shi Pu, Qun Wang and Xin-Nian Wang, Phys. Rev. Lett. {\bf 110}, 262301 (2013).
%
\bibitem{Loganayagam:2012}
R. Loganayagam and P. Surowka, J. High Energy Phs. {\bf 04}, 079 (2012).\\
arXiv:1201.2812[hep-th]
%
\bibitem{Xiao:2005}
D. Xiao, J. Shi and Q. Niu, Phys. Rev. Lett. {\bf 95}, 137204 (2005).
%
\bibitem{Berry}
  M.~V.~Berry, Proc.~R.~Soc.~Lond. A {\bf 392}, 45 (1984)
%
\bibitem{adler69}
S. Adler, Phys. Rev. {\bf 177}, 2426 (1969).
\bibitem{bell69}
J.S. Bell and R. Jackiw, Nuovo Cimento A {\bf 60} 4 (1969).
%
\bibitem{Nielsen:1983rb} 
  H.~B.~Nielsen and M.~Ninomiya,
  Phys.\ Lett.\ B {\bf 130}, 389 (1983).
%
\bibitem{vilenkin:80}
A. Vilenkin, Phys. Rev. D {\bf 22}, 3080 (1980).
%
\bibitem{Fukushima:2008xe} 
  K.~Fukushima, D.~E.~Kharzeev, and H.~J.~Warringa,
  Phys.\ Rev.\ D {\bf 78}, 074033 (2008).
%
\bibitem{Kharzeev08}
D. E. Kharzeev, L. D. McLerran and H. J. Warringa, Nucl. Phys. A {\bf 803}, 67 92007).
%
\bibitem{Warringa08}
H. J. Warringa, arXiv:0805.1384[hep-ph].
%
\bibitem{Abelev:2009}
B. I. Abelev {\it et. al.} [STAR Collaboration], Phys. Rev. Lett. {\bf 103}, 251601 (2009).\\
arXiv:0909.1739 [nucl-ex]
\bibitem{Abelev:2010}
B. I. Abelev {\it et. al.} [STAR Collaboration], Phys. Rev. C. {\bf 81}, 054908 (2010).\\
arXiv:0909.1717[nucl-ex].
%
\bibitem{Xiao:2010}
  D.~Xiao, M.-C.~Chang, and Q.~Niu, 
  Rev.\ Mod.\ Phys.\ {\bf 82}, 1959 (2010). 
  [arXiv:0907.2021 [cond-mat.mes-hall]].
%
\bibitem{kim13}
Heon-Jung Kim {\it et. al.}, Phys. Rev. Lett. {\bf 111}, 246603 (2013). 
%
\bibitem{Sasaki01}
K. Sasaki, arXiv:0106190 [cond-mat].
\bibitem{Son:2012zy} 
  D.~T.~Son and N.~Yamamoto,
  Phys.\  Rev.\ D {\bf 87}, 085016 (2013) [arxiv:1210.815].
\bibitem{Manuel:2013}
C. Manuel and J. M. Torres-Rincon, arXiv:1312.1158[hep-ph].
\bibitem{Itoyama:1983}
H. Itoyama and A. H. Mueller, Nucl. Phys. B, {\bf 165}, 349 (1983).
%
\bibitem{Liu:1988}
Y. Li and G. Ni, Phys. Rev. D. {\bf 38}, 3840 (1988).
%
\bibitem{Nicola:1994}
A. Gomez Nicola and R. F. Alvarez-Estrada, Int. J. Mod. Phys. A {\bf 9}, 1423 (1994).
\bibitem{Akamatsu:2013}
Y. Akamatsu and N. Yamamoto, Phys. Rev. Lett. {\bf 111}, 052002 (2013).
\bibitem{Nieves:1989}
J. F. Nieves and P. B. Pal, Phys. Rev. D {\bf 39}, 652 (1989).
\bibitem{Redlich:1985} 

  A.~N.~Redlich and L.~C.~R.~Wijewardhana,
  Phys.\ Rev.\ Lett.\  {\bf 54}, 970 (1985).

\bibitem{Rubakov:1985} 
  V.~A.~Rubakov,
  Prog.\ Theor.\ Phys.\  {\bf 75}, 366 (1986).

\bibitem{Joyce:1997} 
  M.~Joyce and M.~E.~Shaposhnikov,
  Phys.\ Rev.\ Lett.\  {\bf 79}, 1193 (1997).

\bibitem{Liane:2005} 
  M.~Laine,
  JHEP {\bf 0510}, 056 (2005).

\bibitem{Boyarsky:2012} 
  A.~Boyarsky, J.~Frohlich, and O.~Ruchayskiy,
  Phys.\ Rev.\ Lett.\  {\bf 108}, 031301 (2012).
%
\bibitem{Tsitsadze:2009}
L. N. Tsintsadze, Phys. Plasmas, {\bf 16}, 094507 (2009).
\bibitem{Hu:1991}
B. Y. Hu and J. W. Wilkins, Phys. Rev. B {\bf 43}, 14 009 (1991).

\bibitem{Krall}
N. A. Krall and A. W. Trivelpiece, {\it Principles of Plasma Physics}(San Francisco Press, San Francisco, 1986).
%
\bibitem{Weibel:1959}
E.S. Weibel, Phys. Rev. Lett. {\bf2}, 83 (1959).
%
\bibitem{Fried:1959}
B.D. Fried, Phys. Fluids {\bf 2}, 337 (1959).
%
\bibitem{Mro:1988}
S. Mr\'{o}wczynski, Phys. Lett. B {\bf 214}, 587 (1988),
S. Mr\'{o}wczynski, Phys. Lett. B {\bf 314}, 118 (1993), J. Randrup and S.Mr\'{o}wczynski,
arXiv:0303021[nucl-th]
%
\bibitem{Bhatt:1994}
J. R. Bhatt, P. K. Kaw and J. C. Parikh, Pramana- J. Phys. {\bf 43}, 467 (1994) 
%
\bibitem{Arnold:2003}
P. A. Arnold, J. Lenaghan and G. D. Moore,   J. High Energy Phs. {\bf 08}, 002 (2003).
%
\bibitem{Romatschke:2002}
P. Romatschke and R. Venugopalan, Phys. Rev. Lett. {\bf 96}, 062302 (2006)
%
\bibitem{romatschke}
Paul Romatschke and Michael Strickland, Phys. Rev. {\bf D 68}, 036004 (2003).

%
\bibitem{Kobes:1991}
R. Kobes, G. Kunstatter and A. Rebhan, Nucl. phys. B, {\bf 355} 1 (1991).

\bibitem{Nieves:2000}
See for example Neutrino driven beam instability discussed in this reference:\\
J. F. Nieves, Phys. Rev. D {\bf 61}, 113008 (2000).\\
arXiv:0001067[hep-ph]
%
\bibitem{Ng:2006}
J. S. T. Ng and R. J. Nobel, Phys. Rev. Lett. {\bf 96}, 115006 (2006).
%
\bibitem{Manuel:2008}
M. Mannarelli and C. Manuel, Phys. Rev. D {\bf 77}, 054018(2008).
%
\bibitem{Manuel:2006}
C. Manuel and S. Mr\'{o}wczynski, Phys. Rev. D {\bf 74}, 15003 (2006).
%
\bibitem{Pisarski:97}
R. D. Pisarski, arXiv:9710370[hep-ph] 
%
%
\bibitem{Aad:2011}
G. Aad et al.*(ATLAS Collaboration), Phys. Rev. D {\bf 84} 054001 (2011).
\bibitem{Abelev:2006}
B. Abelev et al. (STAR), Phys. Rev. Lett. {\bf 97}, 252001 (2006).

\end{thebibliography}
\end{document}